\documentclass{aa}
\usepackage{epsfig,psfig}
\usepackage{txfonts}

\newcommand{\msun}{\mbox{$M_{\odot}$}}
\newcommand{\Msun}{\mbox{$M_{\odot}$}}
\newcommand{\lsun}{\mbox{$L_{\odot}$}}

\newcommand{\rsun}{\mbox{$R_{\odot}$}}

\newcommand{\teff}{\mbox{$T_{\rm eff}$}}

\newcommand{\vinf}{\mbox{$v_{\infty}$}}

\newcommand{\mdot}{\mbox{$\dot{M}$}}
\newcommand{\Mdot}{\mbox{$\dot{M}$}}

\newcommand{\msunyr}{\mbox{$M_{\odot} {\rm yr}^{-1}$}}

\newcommand{\kms}{km s$^{-1}$}
\newcommand{\degree}{$^{\rm o}$}

\newcommand{\ha}{H$\alpha$}

\newcommand{\bb}{\bibitem[]{bla}}

\begin{document}

\title{On the presence and absence of disks around O-type stars}

\author{Jorick S. Vink\inst{1}, Ben Davies\inst{2,3}, Tim J. Harries\inst{4}, Ren\'e D. Oudmaijer\inst{2}, Nolan R. Walborn\inst{5}}
\offprints{Jorick S. Vink, jsv@arm.ac.uk}

\institute{Armagh Observatory, College Hill, Armagh BT61 9DG, Northern Ireland, UK
          \and
           The School of Physics and Astronomy, EC Stoner Building, The University of Leeds, Leeds LS2 9JT, UK
          \and
Chester F.\ Carlson Center for Imaging Science, Rochester Institute of Technology, 54 Lomb Memorial Drive, Rochester, NY 14623, USA
          \and
          School of Physics, University of Exeter, Exeter EX4 4QL, UK
          \and
            Space Telescope Science Institute, 3700 San Martin Drive, Baltimore, MD 21218, USA}

\titlerunning{On the presence and absense of disks around O stars}
\authorrunning{Jorick S. Vink}

\abstract {As the favoured progenitors of long-duration gamma-ray bursts, massive stars 
may represent our best signposts of individual objects in the early Universe, but special conditions 
seem required to make these bursters. These are thought to originate from the progenitor's rapid rotation 
and associated asymmetry.} 
{To obtain empirical constraints on the interplay between stellar rotation and wind asymmetry, we perform 
linear H$\alpha$ spectropolarimetry on a sample of 18 spectroscopically peculiar massive O stars, 
including OVz, Of?p, Oe, and Onfp stars, supplemented by an earlier sample of 20 O supergiants 
of Harries et al., yielding a total number of 38 O-type stars.
Our study's global aim is to characterize the differences between, and 
similarities amongst, different classes of peculiar O stars and to establish in how far 
they differ from garden-variety O stars.} 
{Our linear (Stokes $QU$) spectropolarimetry data should be regarded 
a geometric counterpart to (Stokes $I$) spectral classification, setting the stage 
for circular (Stokes $V$) polarimetric searches for magnetic fields.}
{Despite their rapid rotation (with $v$sin$i$ up to $\sim$400\,km/s) most O-type stars 
are found to be spherically symmetric, but with notable exceptions amongst specific object classes. 
We divide the peculiar O stars into four distinct categories: 
Group I includes the suspected young zero-age main sequence OVz stars and related weak-winds objects, 
of which the magnetic star $\Theta^1$ Ori C is the most famous member. These objects show no
evidence for significant linear polarization. Group II includes 
the Of?p stars, in which one of its members, HD 191612, was also found 
to be magnetic. These objects show more linear polarization activity than those in group I.
Group III includes the Oe stars, which have been suggested to be the more massive counterparts to 
classical Be stars, and Group IV concerns the Onfp stars. 
Objects from the latter two groups are on the high-end tail of the 
O star rotation distribution and have in the past been claimed to be embedded in disks. 
Here we report the detection of a classical depolarization ``line effect'' 
in the Oe star HD\,45314, but the overall incidence of line effects amongst Oe stars is significantly lower 
(1 out of 6) than amongst Be stars. The chance that the Oe and Be datasets are 
drawn from the same parent population is negligible (with 95\% confidence). This  
implies there is as yet no evidence for a disk hypothesis in Oe stars, providing relevant 
constraints on the physical mechanism that is responsible for the Be phenomenon. 
Finally, we find that 3 out of 4 of the group IV Onfp stars show evidence for complex polarization effects, which are 
likely related to rapid rotation, and we speculate on the evolutionary links to B[e] stars.}{}


\maketitle

\section{Introduction}
\label{s_intro}

Massive stars have a pronounced effect
on their surrounding interstellar medium and on galaxy 
evolution, but despite their key role in our Universe, huge uncertainties remain 
with respect to their formation and early evolution. 
As long as direct imaging of the 
inner regions around young massive stars is beyond the capabilities of current 
imaging instrumentation, the tool of spectropolarimetry may prove the 
most suitable avenue for obtaining information about 
asymmetric geometries -- of relevance if we want to know which massive stars
make long-duration gamma-ray bursts (GRBs).  

For intermediate-mass young Herbig Ae/Be stars, \ha\ 
spectropolarimetry has been very successful showing
that Herbig Be stars with masses up to $\simeq$10-15 \msun\ are 
embedded in disks on the smallest spatial scales 
(Vink et al. 2002). For the more massive O-type stars, the study of 
their youthful phase is more challenging due to the shorter evolutionary timescales 
as well as the obscuration of the O star(-disk) system by their natal 
dusty cocoons at their earliest times (e.g. Zinnecker \& Yorke 2007). 
To establish the evolutionary links in the early lives of O stars from their 
pre-main sequence, to the zero-age main-sequence (ZAMS), to post main-sequence 
phases, one needs to characterize the differences between and similarities amongst their
class. 

O star {\it spectral} classification has been performed very successfully 
over the last four decades. Important classes in this respect are the so-called
``Oe'' and ``Onfp'' (Walborn 1973; also referred to as ``Oef'' Conti \& Frost 1974) stars. 
The double-peaked Balmer emission-line profiles of Oe stars and the He {\sc ii} $\lambda$4686 emission-line profile
of Onfp stars have been interpreted as signalling ionized 
disks, suggesting they might be higher mass analogues of classical Be stars (Conti \& Leep 1974). 
Evidence for the disk hypothesis has however, even after four decades, not yet been found.

Almost two decades ago, Walborn \& Parker (1992) identified 
a group of potentially young O stars in the Magellanic 
Clouds, the so-called OVz stars, which may be located close to the ZAMS 
(hence the ``z'' designation in their spectral type). 
Walborn (2009) more recently presented a list of suspected ZAMS O stars, identifying 
three categories of ZAMS candidates: (a) a group of Ovz stars which show particularly strong He {\sc ii} $\lambda$4686 absorption, indicative 
of a weak wind and low luminosity 
(Walborn \& Blades 1997), (b) stars with broad and strong hydrogen lines possibly indicating a high gravity, 
and (c) stars with very weak UV wind lines for their spectral types, indicative of a weaker wind 
due to a lower luminosity (see the IUE atlas of Walborn et al. 1985; Heydari-Malayeri et al. 2002). 
We note there is a difference between the phenomenology of ``weak wind stars' which are objects 
with very weak UV wind signatures {\it for their spectral type} and stars that are subject 
to the {\it weak wind problem}, which refers to an issue which seems to appear when comparing 
empirical and theoretical mass-loss rates for O-type stars below a luminosity of log ($L/\lsun$) $\simeq$5.2 
(Martins et al. 2005b, Mokiem et al. 2007, Puls et al. 2008, Marcolino et al. 2009).

Next to spectral Stokes I classification, \ha\ spectropolarimetry adds geometrical 
information, as the linear polarized contribution of starlight has a more selective 
origin than total light.  
In its 
simplest form, the technique is based on the expectation that H$\alpha$ line photons arise 
over a larger volume than stellar continuum photons, such that line photons undergo 
fewer scatterings off the circumstellar disk than continuum photons, and the emission-line 
flux becomes less polarized than the continuum. This results in a smooth polarization 
variation across the H$\alpha$ line profile: the ``line effect''. 
The high incidence of these line effects amongst classical 
Be stars (Vink et al. 2002 counted 26 out of 44 from Poeckert \& Marlborough 1976, i.e. $\sim$55\%) 
indicated that the envelopes of classical Be stars are not spherically symmetric.  
These findings are now taken as compelling evidence that classical 
Be stars are embedded in circumstellar disks (e.g. Waters \& Marlborough 1992, 
Quirrenbach et al. 1997, Oudmaijer et al. 2005).  

In a similar vein, to characterize the early lives of massive O stars 
\ha\ spectropolarimetry may provide an appropriate technique to explore 
source asymmetries, and to search the immediate ionized regions 
for disk signatures.
Harries et al. (2002) performed \ha\ spectropolarimetry on a sample of 20 
mostly normal O supergiants, and found that in the majority of their sample 
line effects were absent (75\%), indicating that despite their rapid rotation 
the environments around O-type stars are {\it generally} spherically symmetric.   
Here we present a linear \ha\ spectropolarimetry survey of 18 peculiar 
O-type stars. This supplements the sample of Harries et al. which consisted of 
16 normal and 4 peculiar Onfp and Of?p stars, yielding a total sample size of 38, 
including 16 normal and 22 peculiar O-type stars.  

We divide the peculiar objects into four groups. ({\sc i})~the suspected young 
zero-age main sequence (ZAMS) stars (OVz and possibly related objects), 
({\sc ii})~the peculiar Of?p stars (with optical carbon {\sc iii} emission lines as strong as the nitrogen {\sc iii} 
lines), for which one of its members HD\,191612 was recently found to be magnetic (Donati et al. 2006).
({\sc iii})~the Oe stars -- the supposedly hotter counterparts of classical Be stars, and finally: 
({\sc iv})~the rapidly rotating Onfp stars (``nfp'' roughly indicating rapid rotation ``n'', strong wind 
emission ``f'' and spectral peculiarity ``p''). 

The goal of our study is to establish the relationships 
between the four groups of peculiar objects and to compare their linear 
polarization properties to those of normal O-type stars. In particular, we  
study the circumstellar environments for signs of youth and circumstellar geometries. 
Making these distinctions is expected to lead to a better 
understanding of both the early evolutionary path of massive stars towards the main sequence, 
as well as the subsequent rotational evolution of O stars beyond the ZAMS.
This is particularly relevant with respect to the GRB phenomenon as special 
circumstances seem required to make them (e.g. Podsiadlowski et al. 2004). 
These are likely associated with the progenitor's rapid rotation, which 
could be obtained in both single (Yoon \& Langer 2005, Hirschi et al. 2005, Woosley \& Heger 2006) 
and binary scenarios (e.g. Cantiello et al. 2007, Wolf \& Podsiadlowski 2007).  
The {\it direct} progenitors of GRBs are believed to be Wolf-Rayet stars, 
for which asymmetry signatures have been found in $\sim$15-20\% of them 
using the tool of linear spectropolarimetry (Harries et al. 1998, Vink 2007). 
This might be accounted for with a scenario in which just the most rapidly 
rotating WR stars produce asymmetries. It seems only natural to assume 
that their rapidly rotating precursor O stars might show line effects 
too.

The paper is organized as follows. 
In Sect.~\ref{s_obs} we briefly describe the observations, data reduction, and analysis
of the linear polarization data, followed by a description of the resulting H$\alpha$ 
line Stokes $I$ and $QU$ linear polarization profiles for the 4 different subgroups of 
peculiar O stars (in Sect.~\ref{s_res}). In Sect.~\ref{s_disc}, we discuss the constraints
these observations provide on the circumstellar environments of the four groups of O stars. We summarize 
in Sect.~\ref{s_sum}.

\section{Observations, data reduction, and methodology}
\label{s_obs}

The linear spectropolarimetry data were 
obtained during two runs in Feb and Aug 2003 using the  
RGO spectrograph of the 4-metre Anglo-Australian Telescope (AAT). Two additional
stars were observed using the ISIS spectrograph 
on the 4-metre William Herschel telescope (WHT) in 2003.
The observations and data reduction of the AAT data is extensively
described in Davies et al. (2005) whilst the same is done for the  
WHT data in Vink et al. (2005b). For the young star $\Theta^1$ Ori C the data
are supplemented by three sets of WHT data from 1995/1996 (see Table~1), for 
which the data reduction procedure is described in Oudmaijer \& Drew (1999).

To analyze the linearly polarized component in the spectra, 
the spectrographs were equipped with the appropriate polarization optics, consisting 
of a rotating half-wave plate and a calcite block to rotate 
and separate the light into two perpendicularly polarized light waves.  
For each exposure, four spectra are recorded: the ordinary (O) and extra-ordinary (E) 
rays of both the target and the sky.  
One complete observation set consists of a series of four 
exposures at half-wave plate position angles of 
0\degree, 45\degree, 22.5\degree, and 67.5\degree~to obtain the linear 
Stokes parameters $Q$ and $U$. 
Polarization and zero-polarization standards were observed 
regularly, revealing an intrinsic instrumental 
polarization of the order of 0.1 per cent. We did not
attempt to correct for this, as our main aim is 
to investigate \ha\ polarimetric signatures relative to the continuum.

The E and O ray data were imported into the polarimetry package 
{\sc ccd2pol}, incorporated in the {\sc figaro} software package ({\sc starlink}). 
The Stokes parameters $Q$ and $U$ were determined, leading
to the percentage linear polarization $P$ and its position angle $\theta$ in the 
following way:

\begin{equation}
P~=~\sqrt{(Q^2 + U^2)}
\end{equation}
\begin{equation}
\theta~=~\frac{1}{2}~\arctan(\frac{U}{Q})
\end{equation}
We note that a position angle (PA) of 0\degree, i.e. North-South on the sky, is represented by
a vector that lies parallel to the positive $Q$ axis, whereas a PA of 90\degree\ (i.e. 
East-West) is positioned in the negative $Q$ direction. Positive and negative $U$ axes 
thus correspond to position angles of respectively 45\degree and 135\degree.

The data were subsequently analyzed using {\sc polmap}. 
The achieved (relative) accuracy of the polarization data is in principle only 
limited by photon-statistics and can be very small (typically 0.01 \%). 
However, the quality and the amount 
of data taken on spectropolarimetric standard stars is at present not 
yet sufficient to reach absolute accuracies below 0.1\% (Tinbergen \& Rutten 1997). 

We note that a non-detection would imply the wind is
spherically symmetric on the sky (to within the
detection limit). The detection limit is inversely dependent on the
signal-to-noise ratio (SNR) of the spectrum, and the contrast of the emission line as the
line-emission is depolarizing the (polarized) flux from the continuum. The
detection limit for the maximum intrinsic polarization $\Delta P_{\rm limit}$ can be represented 
by:

\begin{equation}
\Delta P_{\rm limit} (\%) = \frac{100}{SNR} \times \frac{l/c}{l/c-1}
\label{eq:pint}
\end{equation} 

\noindent where $l/c$ refers to the line-to-continuum contrast. 
This detection limit is most useful for objects with strong emission lines, such 
as H$\alpha$ emission in Luminous Blue Variables (Davies et al. 2005), where the emission
completely overwhelms underlying photospheric absorption. 
However, for the O-type stars studied in this paper, the line emission is much weaker (whilst some 
Group {\sc i} objects even show H$\alpha$ in absorption) and 
the detection limit equation, as provided by Eq.~\ref{eq:pint}, looses its meaning.

In general, we aim for an SNR in the continuum of 1000, or more, corresponding
to changes in the amount of linear polarization of 0.1\%, or less. 
This way, we should be able to infer asymmetry degrees in the form of 
equator/pole density ratios, $\rho_{\rm eq}/\rho_{\rm pole}$ of $\sim$1.25, or larger 
(e.g. Harries et al. 1998), with some small additional dependence on the shape and 
inclination of the disk.

Just as we do not correct for instrumental polarization,
no correction for interstellar polarization (ISP) is made either, as 
the ISP only adds a wavelength-independent polarization vector to 
all observed line and continuum wavelengths. 

\section{Results}
\label{s_res}

The linear continuum polarization of O-type stars is thought to be caused by  
scattering of stellar photons off electrons in the circumstellar environment, but 
in addition there may be an interstellar component to the measured level of polarization.  
The continuum (excluding the H$\alpha$ line) polarization data for all our targets 
are summarized in Table~\ref{t_cont} in the form of the 
mean $R$-band percentage polarization and its position angle $\theta$ (columns 4 and 5). 
Polarization variability is commonplace amongst O-type stars (e.g. Hayes 1975, Lupie \& Nordsieck 1987), and 
accordingly we do not look for perfect 
agreement between our continuum polarization measurements and those in 
earlier literature. Nevertheless, the values of the polarization 
quantities \%Pol and $\theta$ are generally consistent with previous 
continuum measurements (see columns 6 and 7). 

Following Harries et al. (2002), we constructed maps of the ISP 
through searching the polarization catalogue 
of Mathewson et al. (1978) for objects lying within 5\degree\ from our targets. 
In some cases, the ISP maps appear both well-ordered and well-populated and in these 
cases column (10) states ``OK''. In case of sightlines where the ISP maps appear disordered
(or sparsely populated) column (10) states ``No''.
In those cases where the match between the measured polarization and the ISP is good, 
the overall polarization is assumed to be dominated by the ISP, and the 
intrinsic polarization will be small. 

\begin{table*}
\caption{Polarization data for four groups of peculiar O-type stars observed during the AAT and WHT runs described in the body text.} 
\label{t_cont}
\begin{tabular}{lllcrcrcrc}
\hline
Name & Spec.Tp & Date & $P_{\rm cont}^{\rm R}$ (\%) & $\theta_{\rm cont}^{\rm R}$ (\degree) & $P^{\rm lit}$ (\%) & $\theta^{\rm lit}$ & $P^{\rm IS}$ (\%) & $\theta^{\rm IS}$ & ISP map\\
\hline
\\
(I)~~{\it OVz and weak wind dwarfs}:\\
\\
\\
$\Theta^1$ Ori C & O7 Vp & 31-12-95 &0.31 $\pm$ 0.01 & 139.9 $\pm$ 0.3 &  &  &  & & \\
                &           & 01-01-96 &0.29 $\pm$ 0.01 & 148.2 $\pm$ 0.8 &  &  & & &  \\
                &           & 31-12-96 &0.29 $\pm$ 0.01 & 144.0 $\pm$ 0.6 &  &  & & &  \\
                &           & 14-02-03 &0.284 $\pm$ 0.005 & 136.3 $\pm$ 0.5 & 0.35 & 151.8 & 0.240 & 88.5 & No\\
$\Theta^2$ Ori A & O9.5 Vpe   & 14-02-03 &1.051 $\pm$ 0.005 & 99.7 $\pm$ 0.1 & 0.702 & 93.1 & 0.221 & 88.5 & No\\
HD 42088 & O6.5 V  & 14-02-03 &  2.431 $\pm$ 0.008 & 177.0 $\pm$ 0.1 & 2.53 & 178 & 2.126 & 168.8 & OK \\
HD 54662  & O6.5 V & 14-02-03 & 0.881 $\pm$ 0.098 & 145.5 $\pm$ 3.8 & 0.83 & 145 & 0.481 & 154.4 & No\\
HD 93128  & O3.5 V((f+)) &  12-02-03 &2.265 $\pm$ 0.014 &  89.0 $\pm$ 0.2 &      &      &       &\\
HD 93129B & O3.5 V((f+)) & 12-02-03 &1.889 $\pm$ 0.063 & 117.5 $\pm$ 1.0 &      &      &       &\\
HD 152590 & O7.5 V & 14-02-03 & 0.998 $\pm$ 0.009 & 29.1 $\pm$ 0.3 & 0.96 & 19.2 & 0.737 & 27.9 & No\\
CPD-58\degree 2611 & O6 V((f))     & 13-02-03 & 2.222 $\pm$ 0.010 &  84.7 $\pm$ 0.1 &      &      &       &\\
CPD-58\degree 2620 & O6.5 V((f))   & 14-02-03 & 1.717 $\pm$ 0.009 &  90.6 $\pm$ 0.1 &      &      &       &\\
\\
\hline
\\
(II)~~{\it Of?p}:\\
\\
\\
HD 148937     & O6.5 f?p  & 12-02-03 &  1.580 $\pm$ 0.020 &  46.9 $\pm$ 0.4 & 1.63 & 50.6 & 1.727 & 36.1 & OK\\
              &           & 15-08-03 &  1.803 $\pm$ 0.014 &  49.3 $\pm$ 0.2 &      &      &       & & \\
\\
\hline
\\
(III)~~{\it Oe stars}:\\
\\
\\
HD 45314 & O9:pe  & 14-02-03 &  1.316 $\pm$ 0.091 & 167.4 $\pm$ 2.0  & 1.38 & 171 & 1.369 & 171.6 & OK \\
HD 46485  & O7 Vn(e) & WHT 03   &  1.887 $\pm$ 0.013 & 26.1 $\pm$ 0.2    & 1.89 & 159.0 & 0.808&165.8& OK\\
HD 60848  & O8 V:pev & WHT 03   &  0.431 $\pm$ 0.007 & 135.7 $\pm$ 0.4& 0.78 & 56  & 0.372 & 27.0  & No\\
HD 120678 & O8 III:nep & 13-02-03 &  1.637 $\pm$ 0.015 & 77.8  $\pm$ 0.3    & 1.7  & 76  & 1.781 & 71.7  & OK\\
Zeta Oph & O9.5 Vnn  & 14-02-03 &  1.572 $\pm$ 0.004 & 127.3 $\pm$ 0.1 & 1.42 & 127 & 0.411 & 76.4  & No\\
HD 155806 & O7.5 V[n]e & 14-08-03 &  0.821 $\pm$ 0.013 & 156.4 $\pm$ 0.5   & 0.86 & 147 & 0.742 & 159.0 & OK\\
\\
\hline
\\
(IV)~~{\it Onfp}:\\
\\
\\
HD 152248 & O7 Ib:(n)(f)p  & 15-08-03 & 0.477 $\pm$ 0.016  & 114.9 $\pm$ 1.0 & 0.66 & 112 & 0.812 & 27.9 & OK\\
\\
\hline
\end{tabular}
\\
\noindent
The spectral types are taken from the O star catalogue of Maiz-Apellaniz et al. (2004) and listed in column 2, with 
the dates of the observations provided in the third column.
The errors in the polarization data (column 4) are of the order of 0.01\% based 
on photon-statistics only. Yet, systematic (external) errors in the 
polarization are estimated to be 0.1\%. 
The errors in the Position Angle $\theta$ (column 5) 
are of order a degree. The literature values of the polarization and PA are taken from the
catalogue of Mathewson et al. (1978) and so are the values for the ISP (columns 8 and 9). The last column (10) 
indicates whether an ISP map is well ordered and well populated (see body text in Sect.~\ref{s_res} for more detailed explanation).
\end{table*}

The H$\alpha$ line shapes of our O-star targets include absorption, emission, 
double-peaked emission, and P~Cygni profiles. 
Plots of the spectropolarimetric data 
are presented in the different panels of Figs.~\ref{f_theta} 
to \ref{f_onfp}. The polarization spectra are
presented as triplots, consisting of Stokes I, $P$, and 
$\theta$.

\subsection{OVz stars}
\label{s_ovz}

\begin{figure*}
\mbox{
\epsfxsize=0.25\textwidth\epsfbox{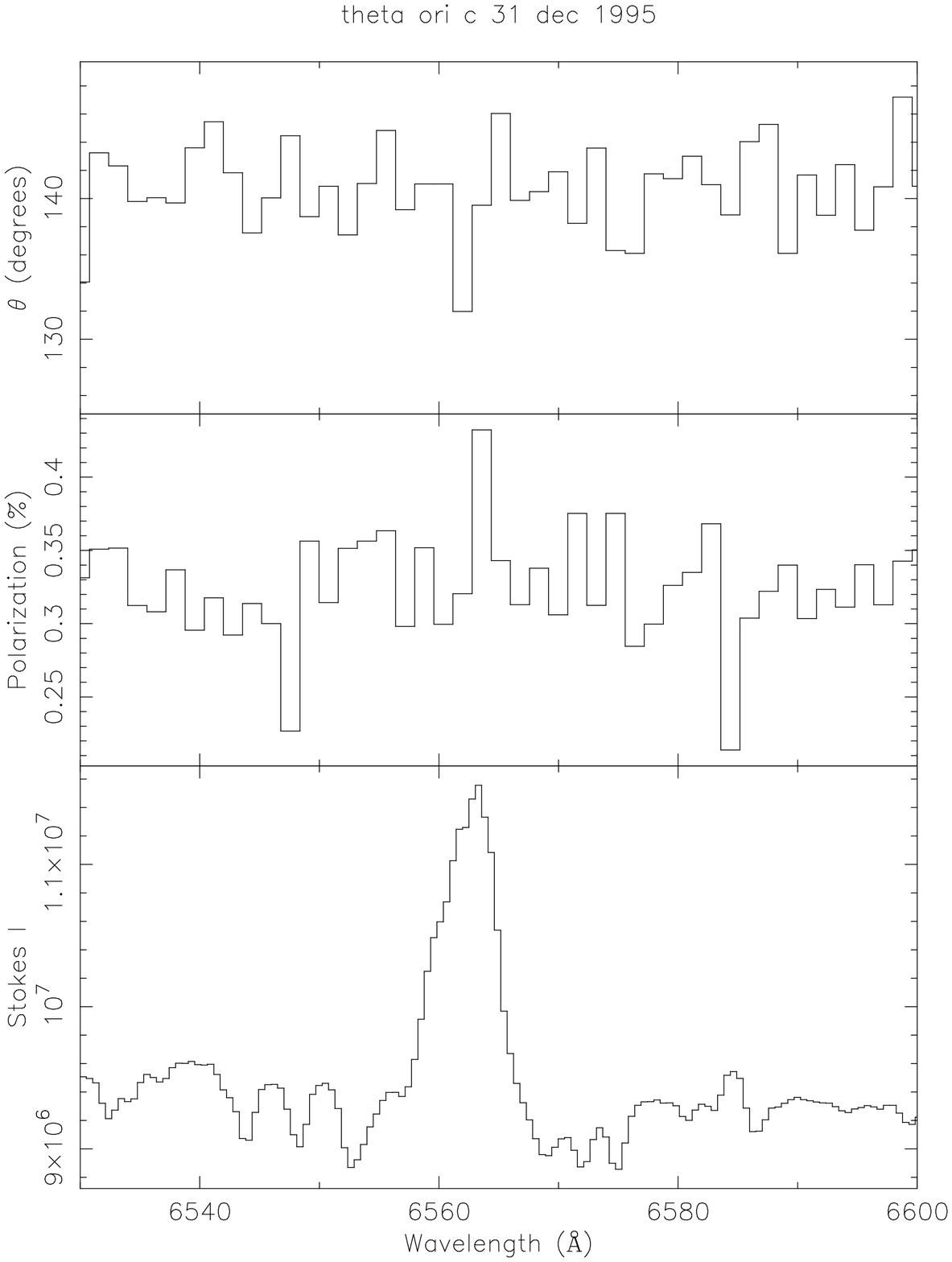}
\epsfxsize=0.25\textwidth\epsfbox{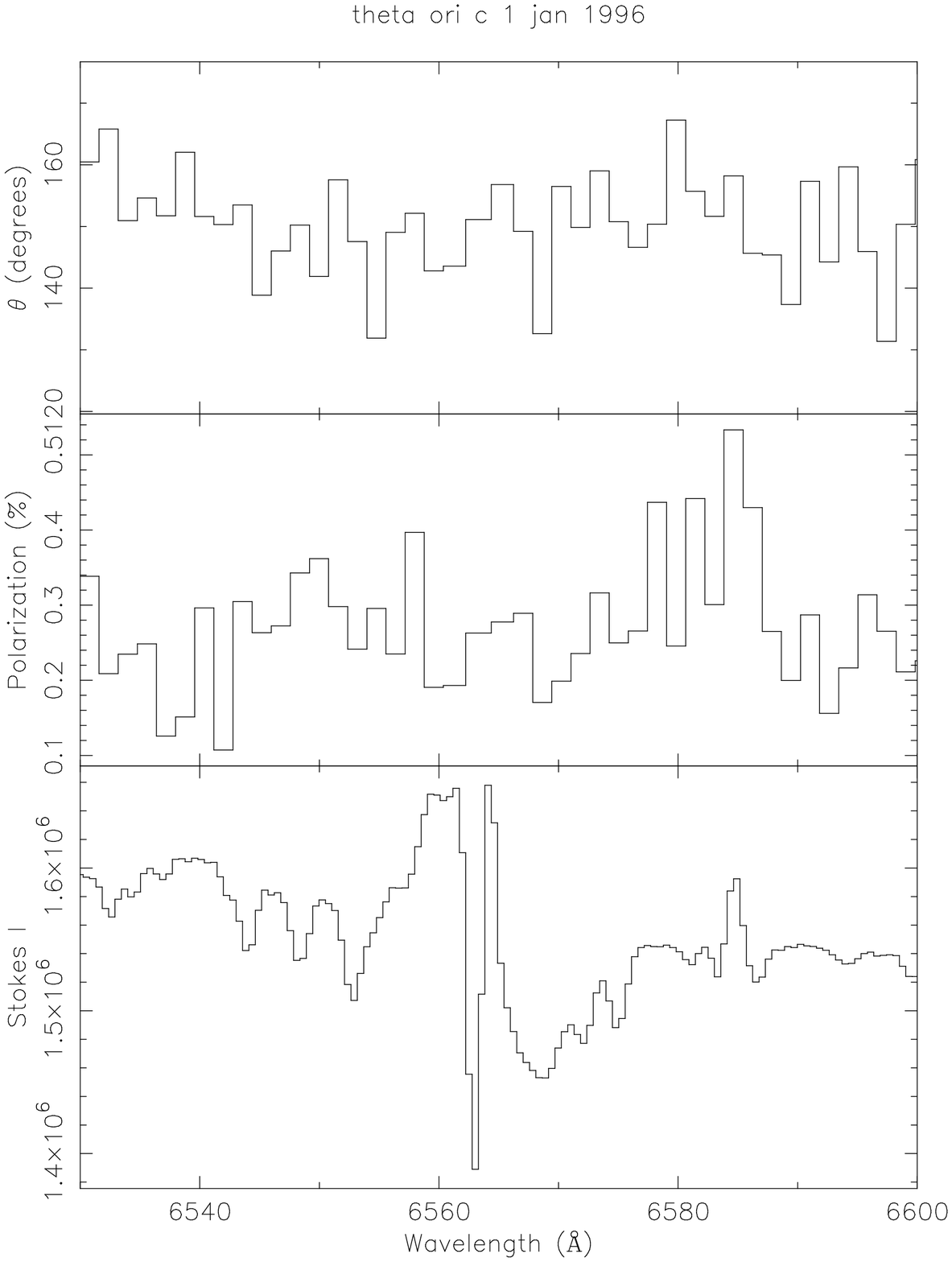}
\epsfxsize=0.25\textwidth\epsfbox{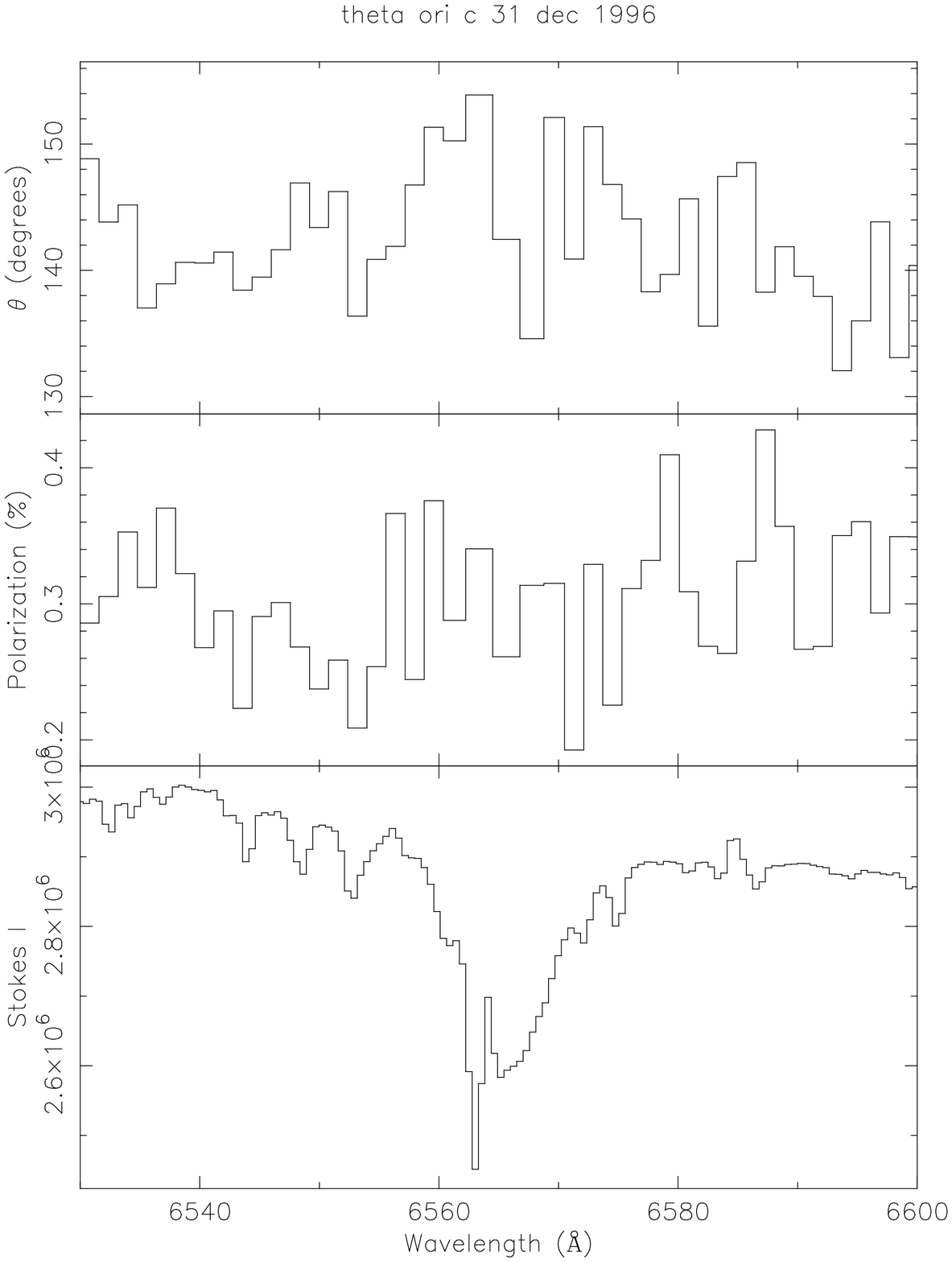}
\epsfxsize=0.25\textwidth\epsfbox{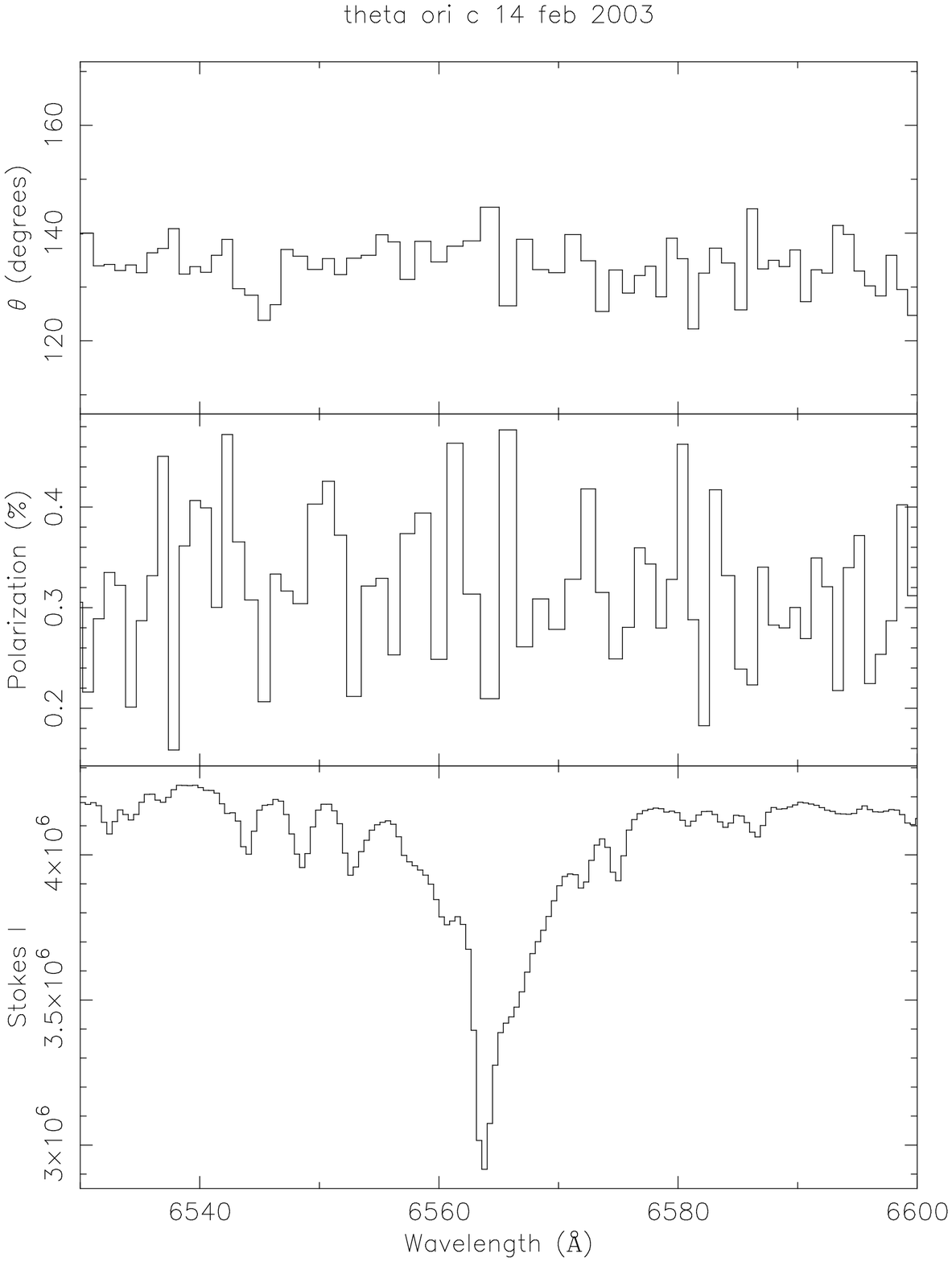}
}
\caption{Polarization spectra of the Group {\sc i} peculiar O star $\Theta^1$ Ori C 
at four differentepochs. 
Stokes I spectra are shown 
in the lowest panels of the triplots, 
the levels of \%Pol in the middle panel, 
whilst the PAs ($\theta$; see Eq.~2) are 
plotted in the upper panels. The data are 
rebinned such that the 1$\sigma$ error in the polarization
corresponds to 0.05\% or 0.1\% as calculated from photon statistics. 
The narrow features that appear in three of the Stokes $I$ line profiles are due to imperfect 
nebular subtraction.}
\label{f_theta}
\end{figure*}

\begin{figure*}
\begin{center}
\mbox{\epsfxsize=0.25\textwidth\epsfbox{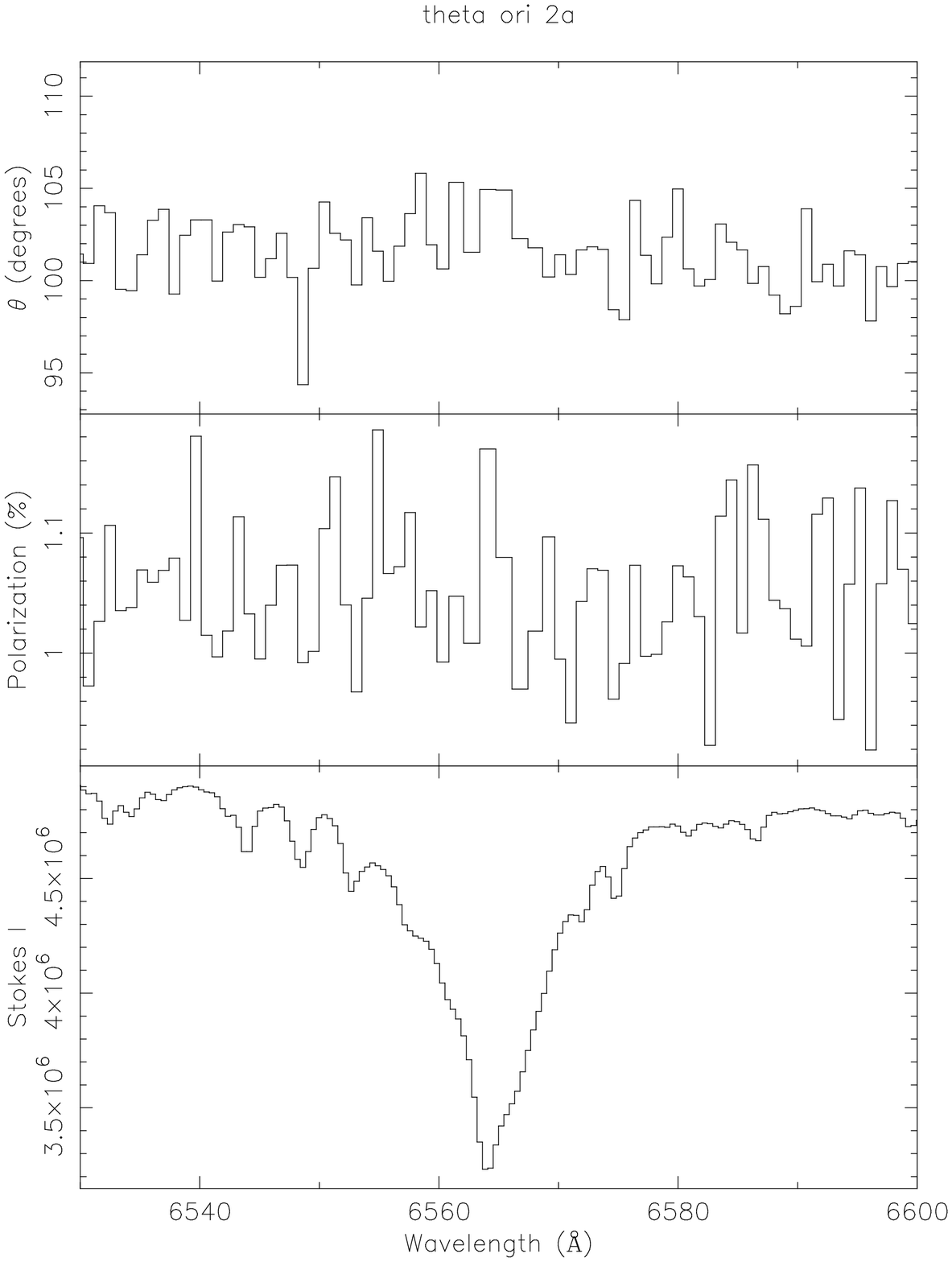}
\epsfxsize=0.25\textwidth\epsfbox{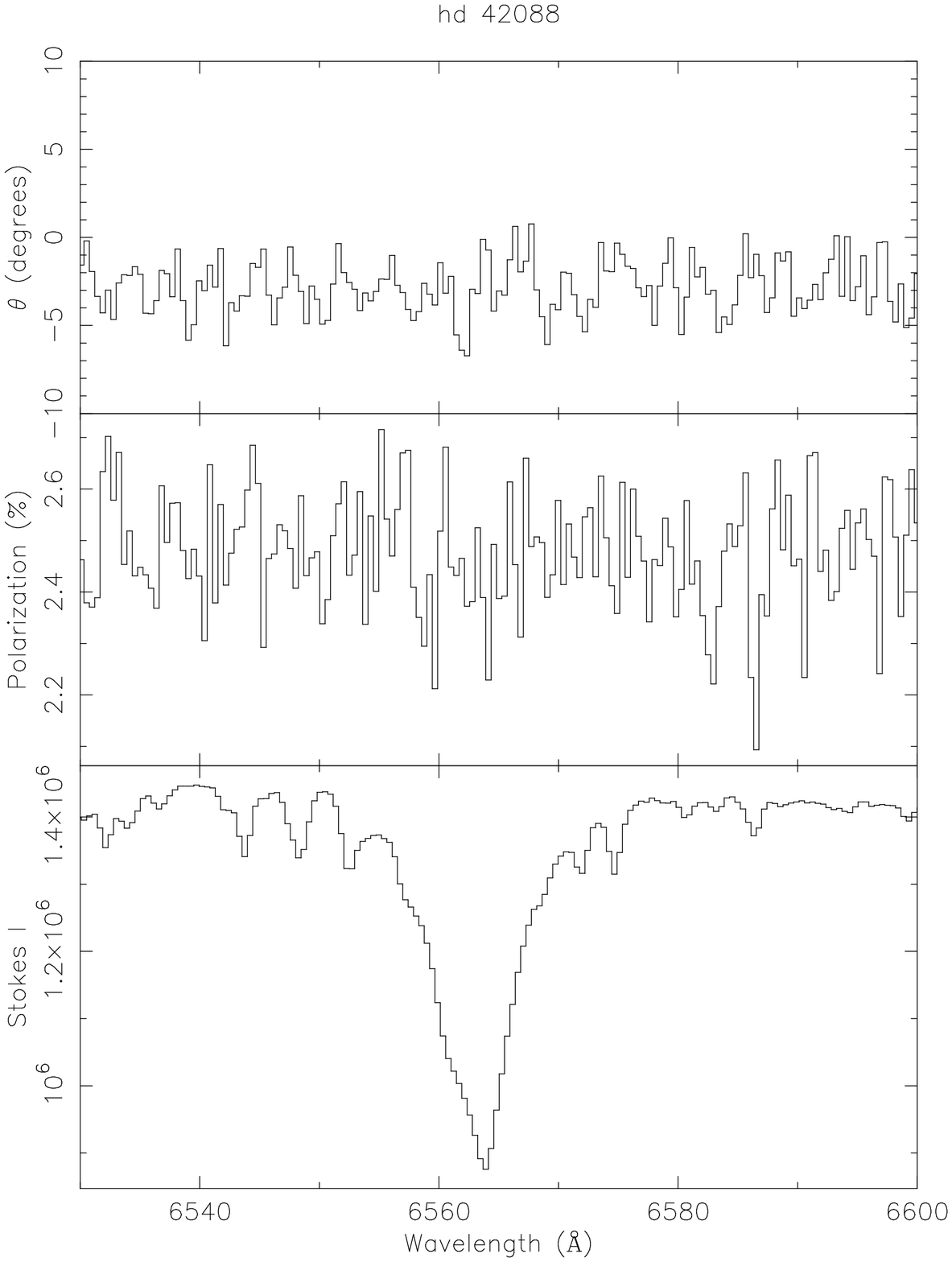}
\epsfxsize=0.25\textwidth\epsfbox{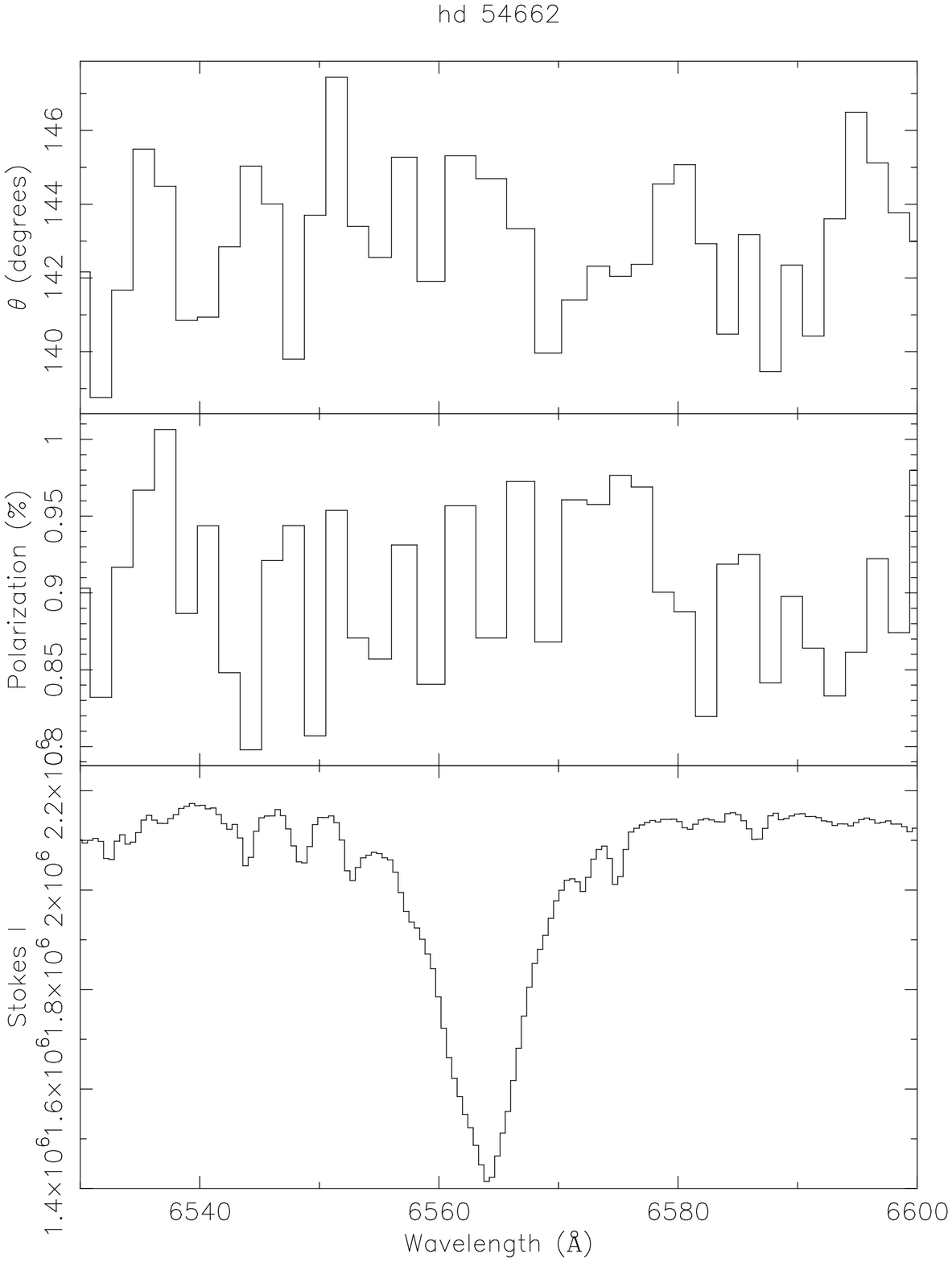}
\epsfxsize=0.25\textwidth\epsfbox{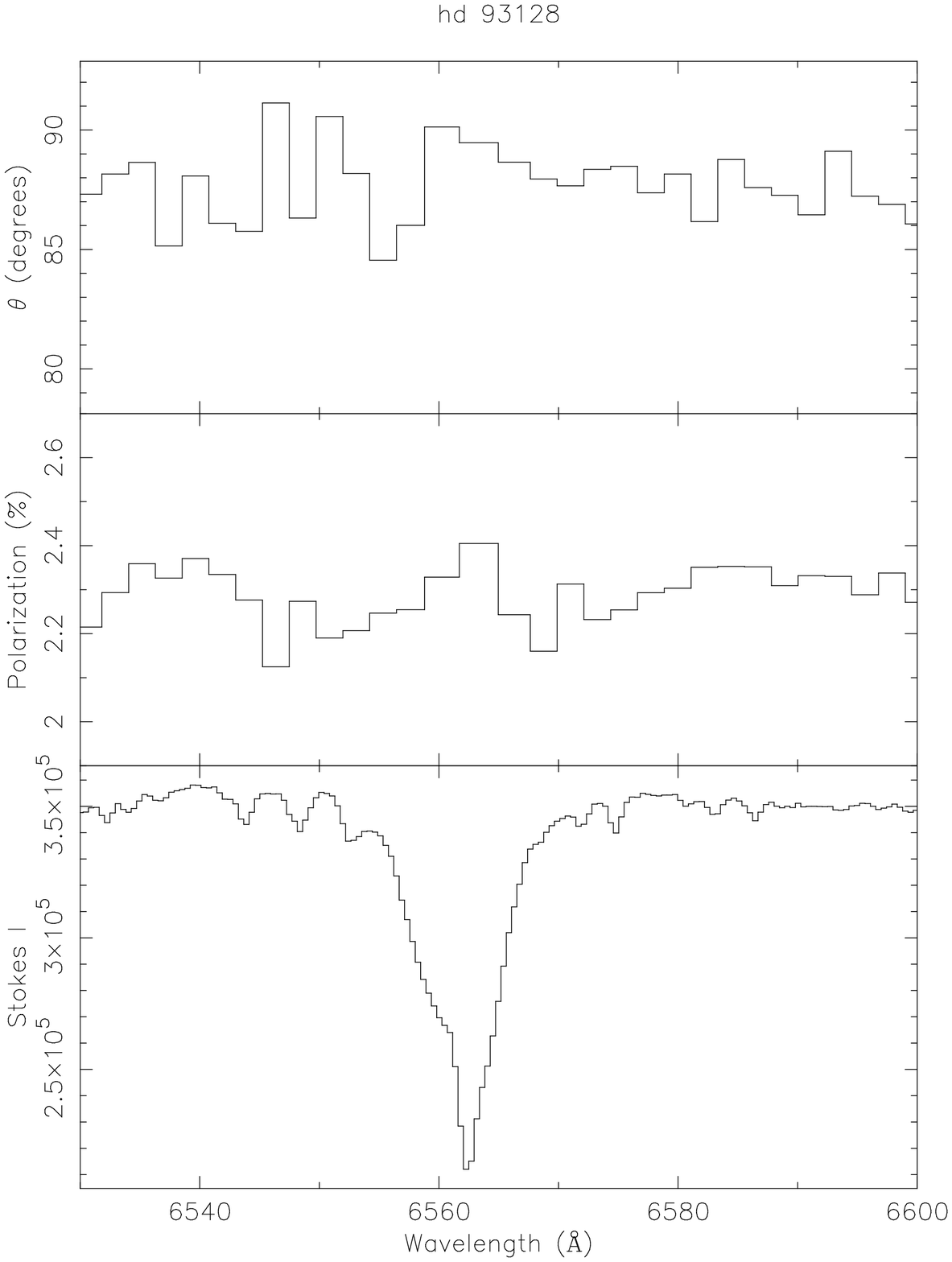}}
\mbox{
\epsfxsize=0.25\textwidth\epsfbox{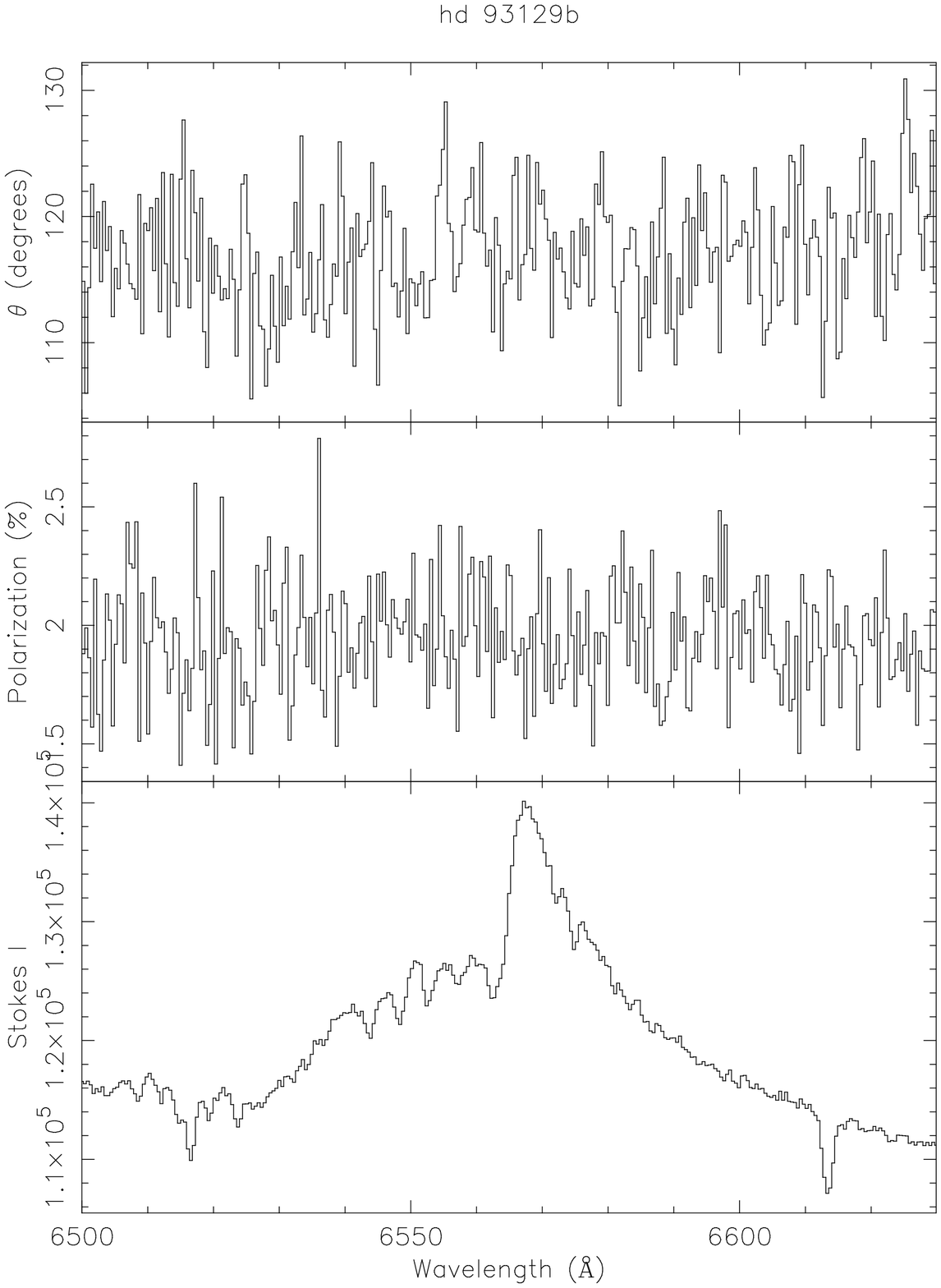}
\epsfxsize=0.25\textwidth\epsfbox{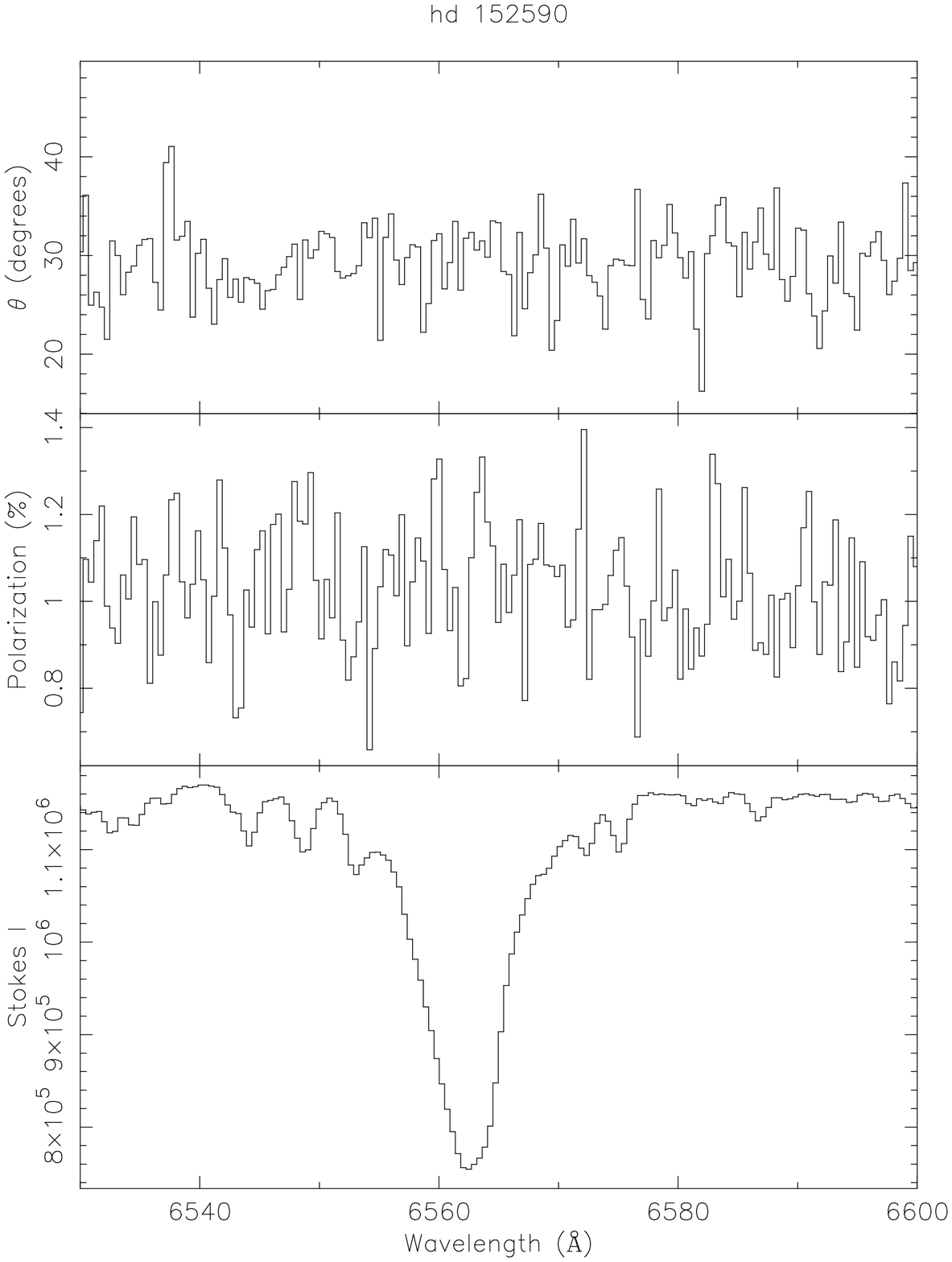}
\epsfxsize=0.25\textwidth\epsfbox{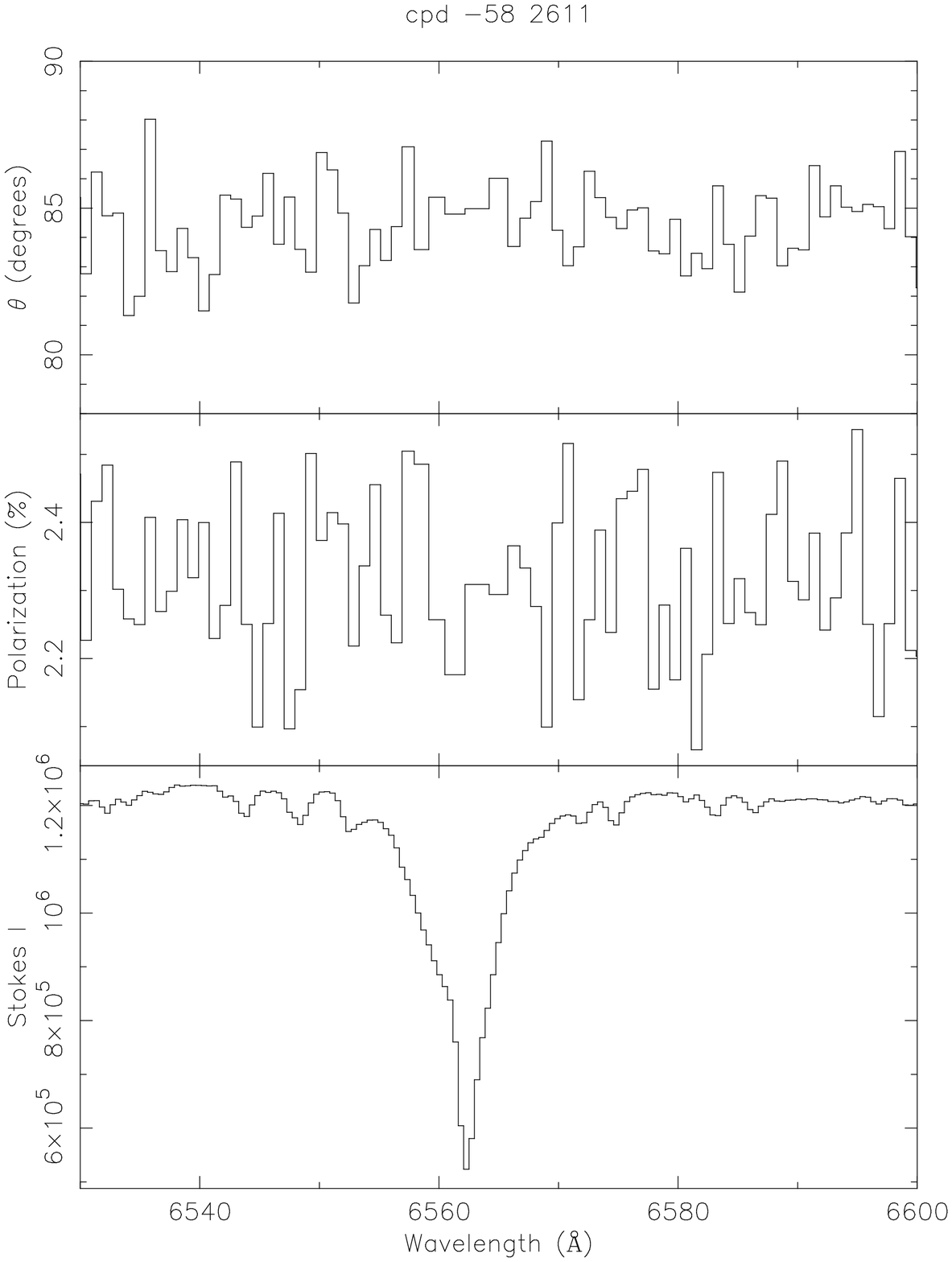}
\epsfxsize=0.25\textwidth\epsfbox{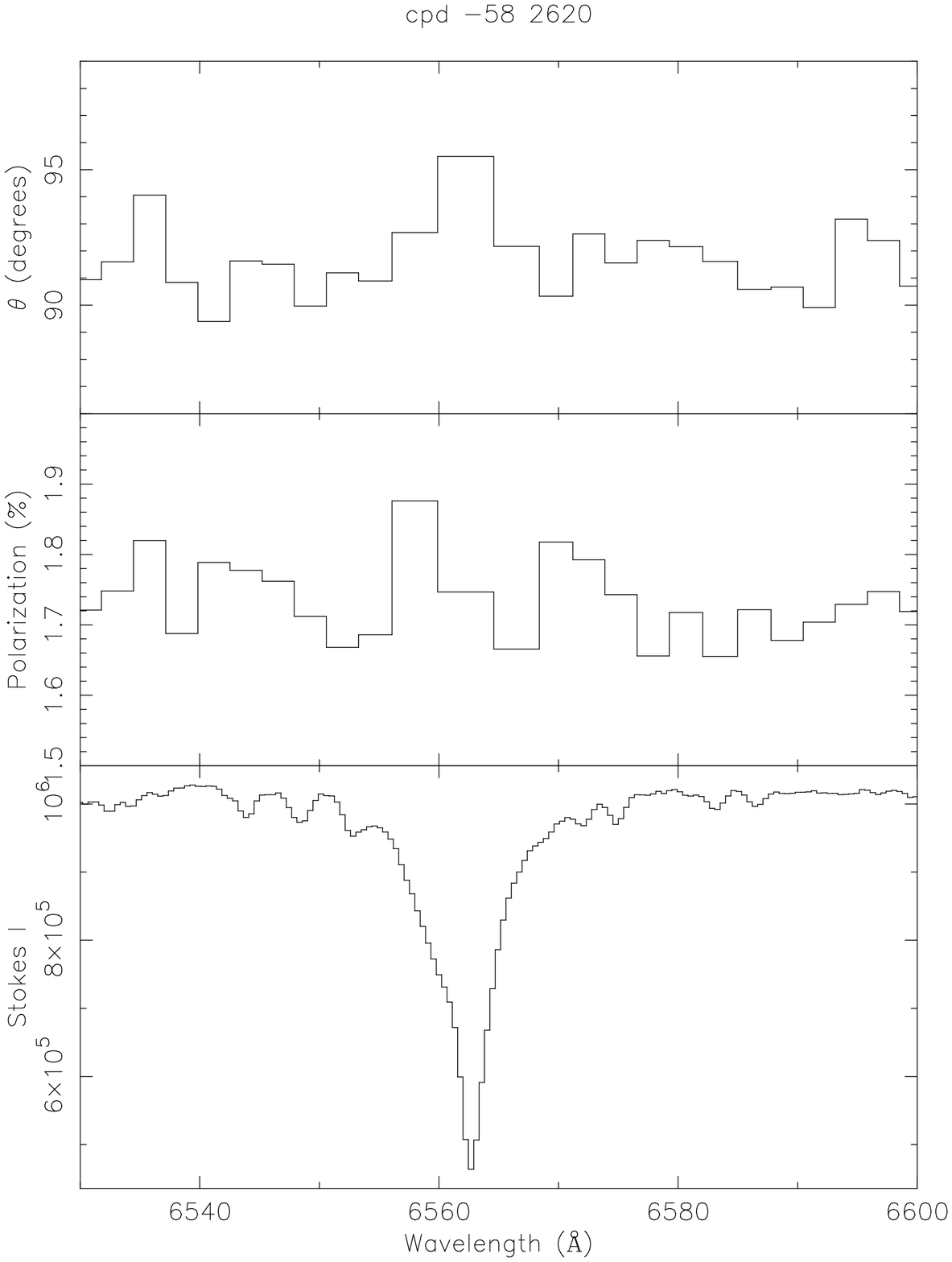}}
\caption{Polarization spectra of the remaining Group {\sc i} ZAMS candidate O stars. Note that the spectrum
of HD93129B is dominated by its 1.5 mag brighter companion HD93129A at only 3 arcsec away.}
\label{f_ovz}
\end{center}
\end{figure*}

The membership criteria for OVz and related ZAMS candidates have been
summarized in Sect.~\ref{s_intro} and Walborn (2009).

The H$\alpha$ line polarimetry data for the 9 ZAMS candidate O-type stars 
are very similar in the sense that H$\alpha$ is mostly in absorption (see Fig.\ref{f_ovz}). 
In none of them do we find evidence for a variation in 
the polarization percentage or position angle across the line. 
In order to address the question of whether this implies 
the objects are intrinsically unpolarized, we compare the continuum polarization data 
to previous measurements, as well as to the ISP.

We start off with the two Orion ZAMS candidate objects $\Theta^1$ Ori C and $\Theta^2$ Ori A. 
As these two objects are members of the Orion nebula cluster (ONC) star
forming region, it should be no surprise that the ISP maps for 
the objects appear disordered. The same is true for the four OVz stars in 
the Carina star forming region, c.f. HD\,93128, HD\,93129B, CPD-58\degree 2611, and CPD-58\degree 2620, but
for these four objects we have no information regarding previous linear polarization data.

In our AAT data of $\Theta^1$ Ori C and $\Theta^2$ Ori A, the H$\alpha$ lines are in absorption, and 
although this does not necessarily imply there cannot be any line effects,
it is still the case that line effects are easier to detect across H$\alpha$ when the line is seen in 
emission, which happens on a regular basis for the oblique rotator 
$\Theta^1$ Ori C (Stahl et al. 1996, Donati et al. 2002, Gagne et al. 2005). 
Multiple observations of $\Theta^1$ Ori C over four different epochs are plotted in Fig.~\ref{f_theta}.
The Stokes I H$\alpha$ line is seen to vary from one night to the next from a pure emission line on 
31 Dec 1995 to an inverse P\,Cygni profile on 1 Jan 1996. In our data from 31 Dec 1996 and 
14 Feb 2003 the line displays a pure absorption profile. We note that the very narrow features 
that appear close to line centre are simply the result of imperfect nebular subtraction.
Despite the large differences in the character of the H$\alpha$ line, the linear polarization properties  
of $\Theta^1$ Ori C (see Table 1) do not appear to vary significantly.

Given that our continuum polarization measurement for the other Orion object
$\Theta^2$ Ori A appears to have changed slightly with respect to earlier
measurements (column 7), this might indicate that the spectroscopic binary 
$\Theta^2$ Ori A is variable in its intrinsic linear
polarization properties, which
might be consistent with wind-wind interaction or
polarization variations due to a magnetically confined wind, such as in the Bp object
$\sigma$Ori E (Townsend \& Cohen, in prep.).

For the remaining three candidate ZAMS O-type stars, cf. HD\,42088, HD\,54662, and HD\,152590, our 
PA values are consistent with earlier measurements and the ISP, and it is probably 
safe to state that the measured polarization for these objects is predominately of interstellar origin, with
only an exceedingly small intrinsic polarization component. 

\subsection{Of?p stars}
\label{s_of?p}

\begin{figure*}
\mbox{
\epsfxsize=0.33\textwidth\epsfbox{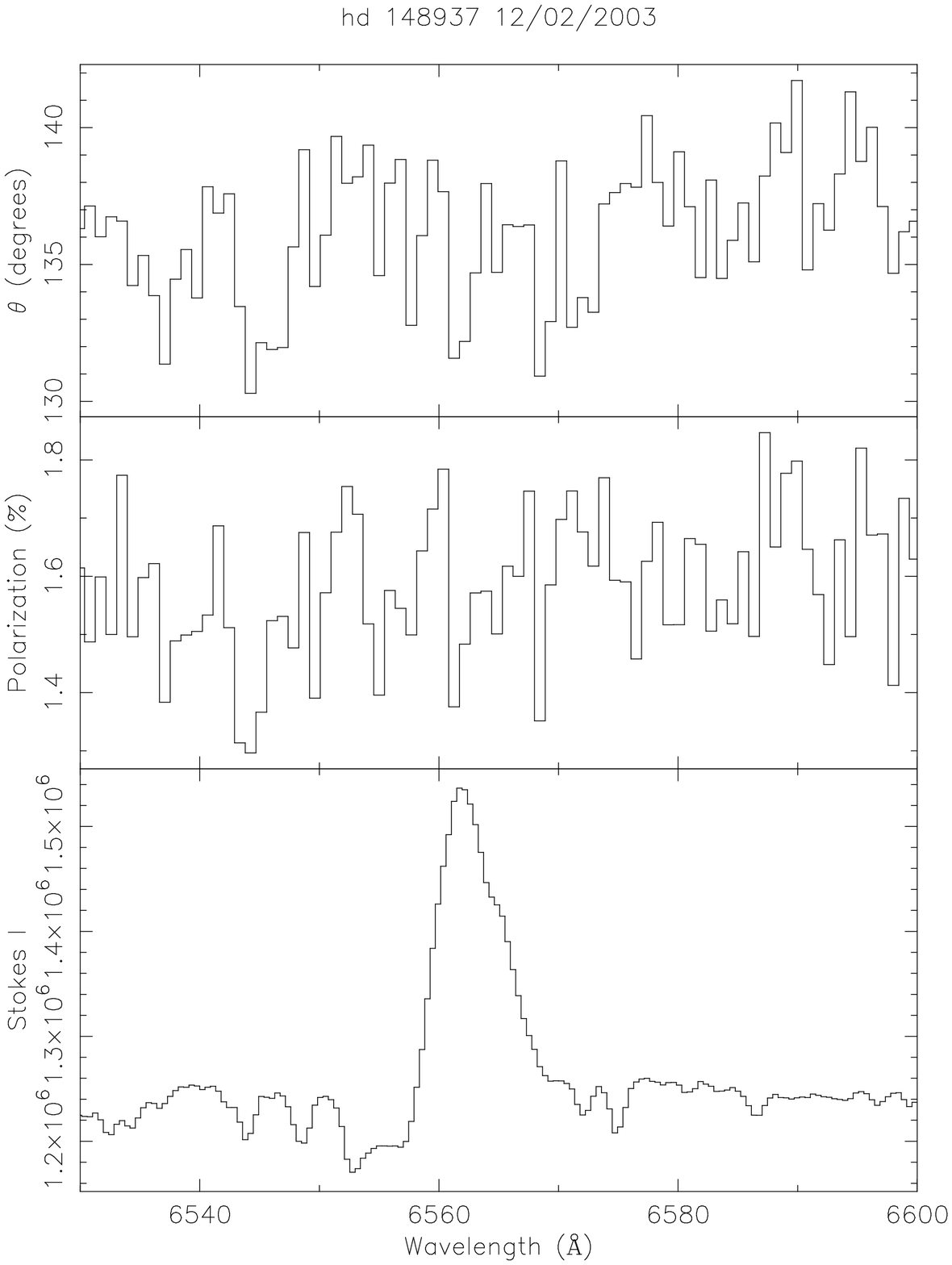}
\epsfxsize=0.33\textwidth\epsfbox{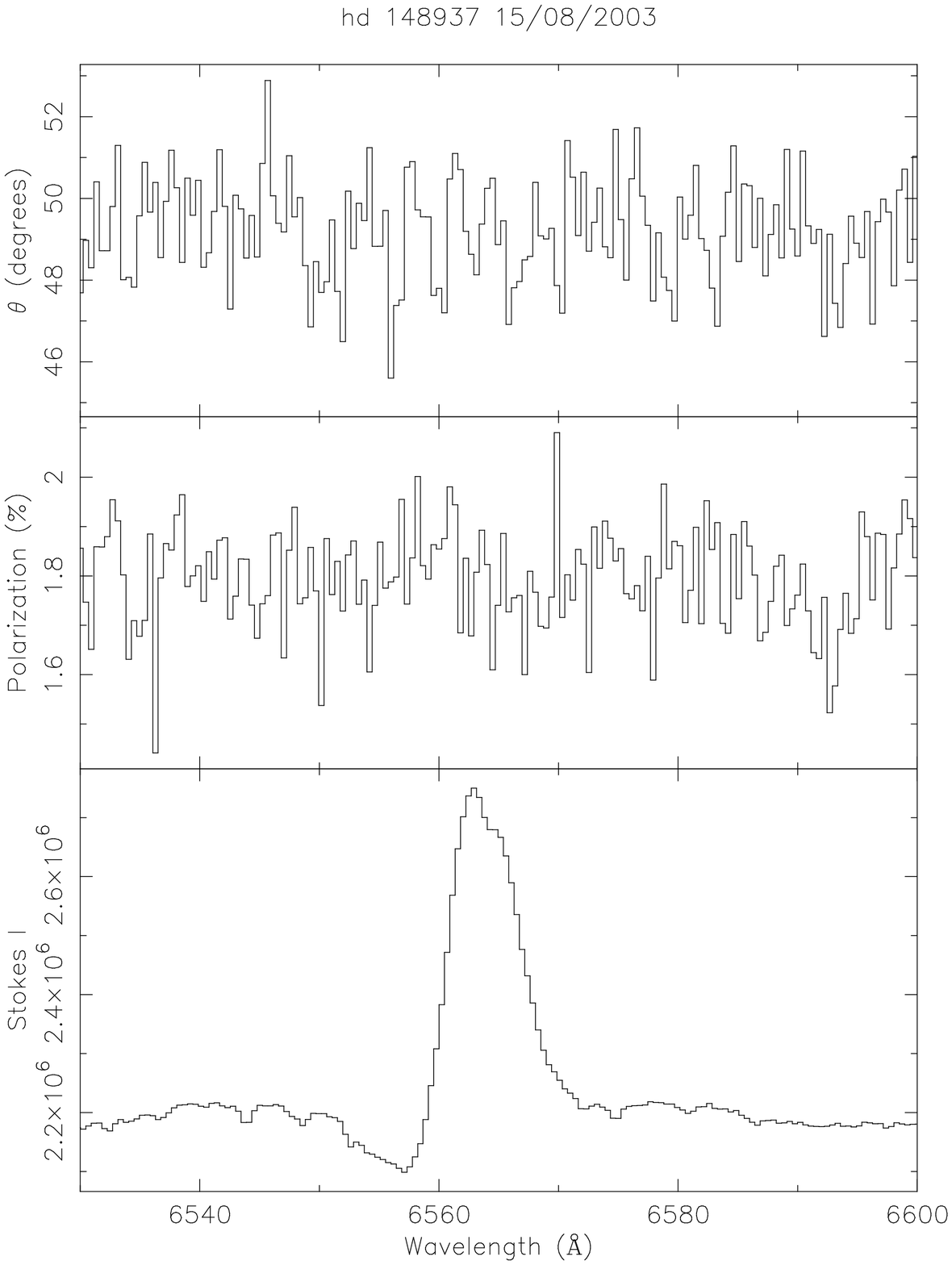}
\epsfxsize=0.33\textwidth\epsfbox{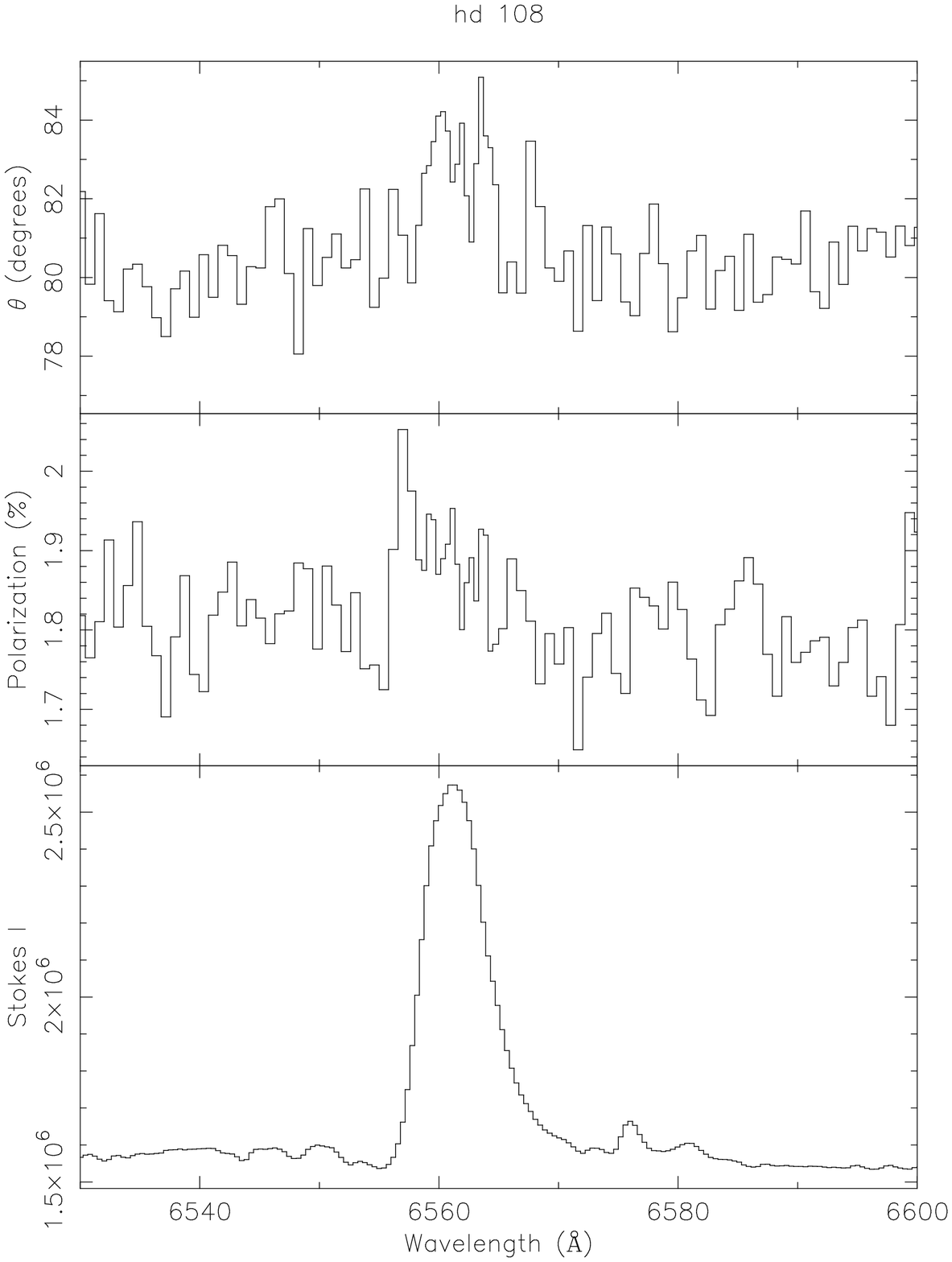}
}
\caption{
Polarization spectra of the Group {\sc ii} Of?p stars HD\,148937
(during both the Feb and Aug 2003 runs) and HD\,108 (from Harries et al. 2002). 
The data are 
rebinned such that the 1$\sigma$ error in the polarization
corresponds to 0.05\% or 0.1\% as calculated from photon statistics.
\label{f_of?p}
}
\end{figure*}

The mysterious Of?p phenomenon is referred to when the carbon {\sc iii}
optical emission at $\sim$4650 \AA\ is equally strong as
that of the nitrogen {\sc iii} emission at 4640\AA\ (see e.g. Maiz-Apellaniz et al. 2004).
Only three such objects are known in the Milky Way. These 
are HD\,191612, which is magnetic (Donati et al. 2006), 
HD\,108, and HD\,148937.

Linear spectropolarimetry data on two different epochs in Feb and Aug 2003 
for HD\,148937 are presented in Fig.~\ref{f_of?p}, alongside data for HD\,108 from
Harries et al. (2002).  
Although both epochs of HD\,148937 show a P Cygni profile in H$\alpha$, there 
are notable changes in the absorptive part of the P~Cygni 
profile. Neither epoch however shows a significant change in the 
polarization character across the H$\alpha$. 
The continuum PA of the two datasets has varied only slightly, from $\theta$ $=$ 49.9\degree\ $\pm$ 0.4\degree\ in Feb 
2003 to $\theta$ $=$ 49.3\degree\ $\pm$ 0.2\degree\ in Aug 2003, i.e. the same within the errors, 
although the level of polarization may have increased from $P$ $=$ 1.58\% to 
1.80\% over the same time interval, but as the systematic error for these kind of data 
is of order 0.1\% (see Sect.~\ref{s_obs}), this only constitutes 
a 2$\sigma$ effect.

\subsection{Oe stars}
\label{s_oe}

Oe stars are defined as near main sequence O-type stars with their Balmer
lines in emission, possibly representing the higher mass analogues of classical Be 
stars (Walborn 1973, Negueruela et al. 2004).

\begin{figure*}
\mbox{
\epsfxsize=0.25\textwidth\epsfbox{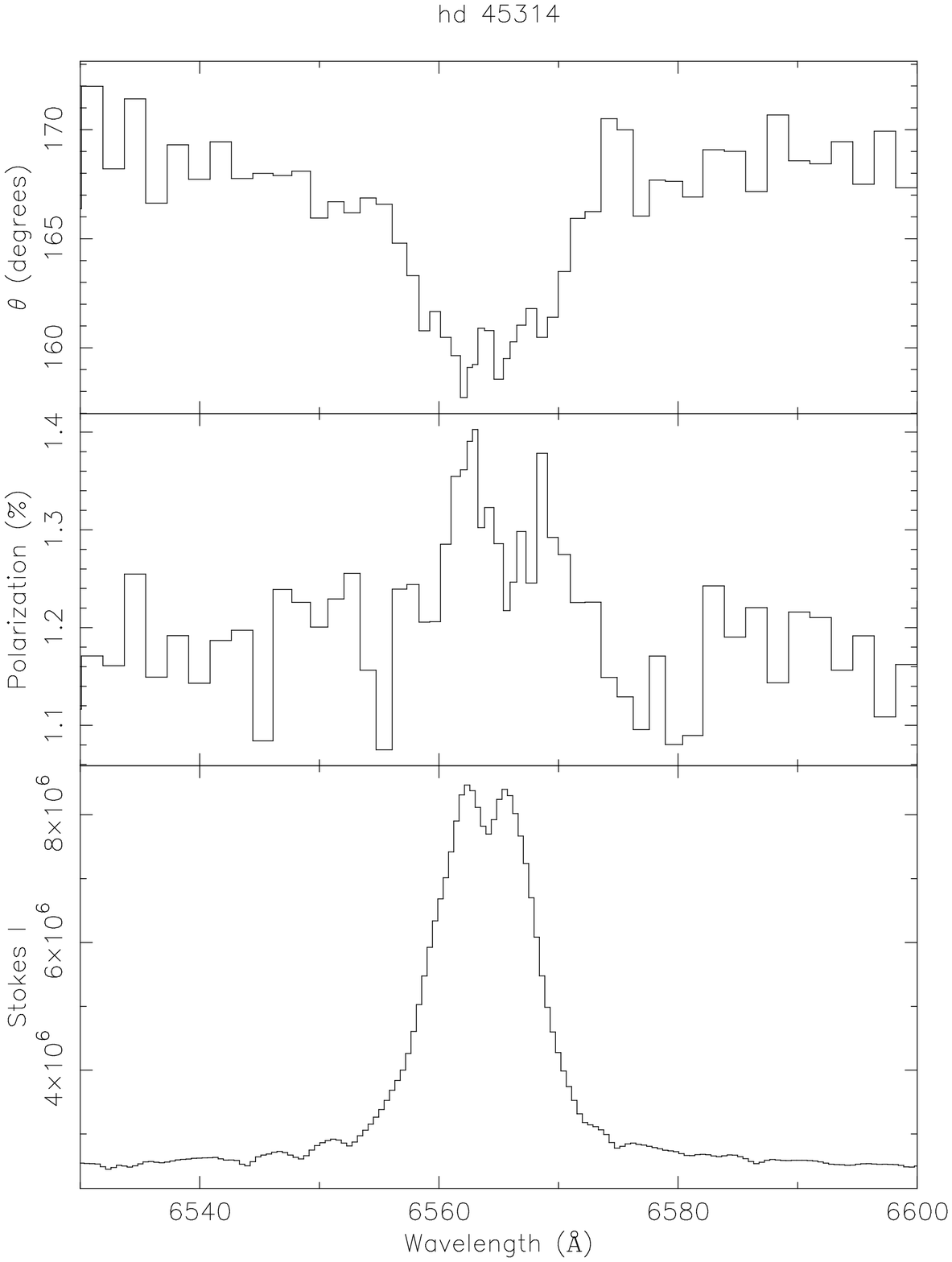}
\epsfxsize=0.25\textwidth\epsfbox{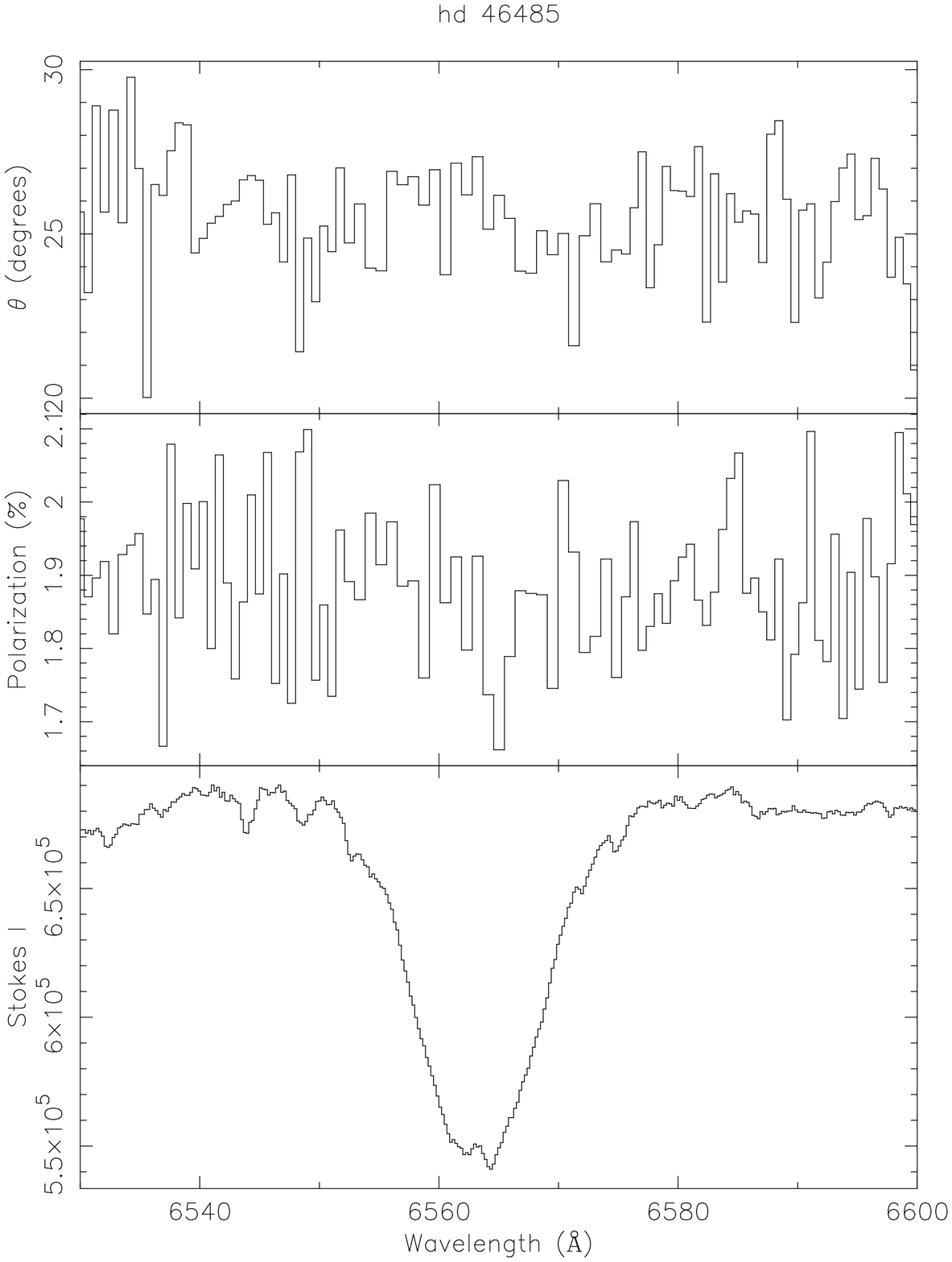}
\epsfxsize=0.25\textwidth\epsfbox{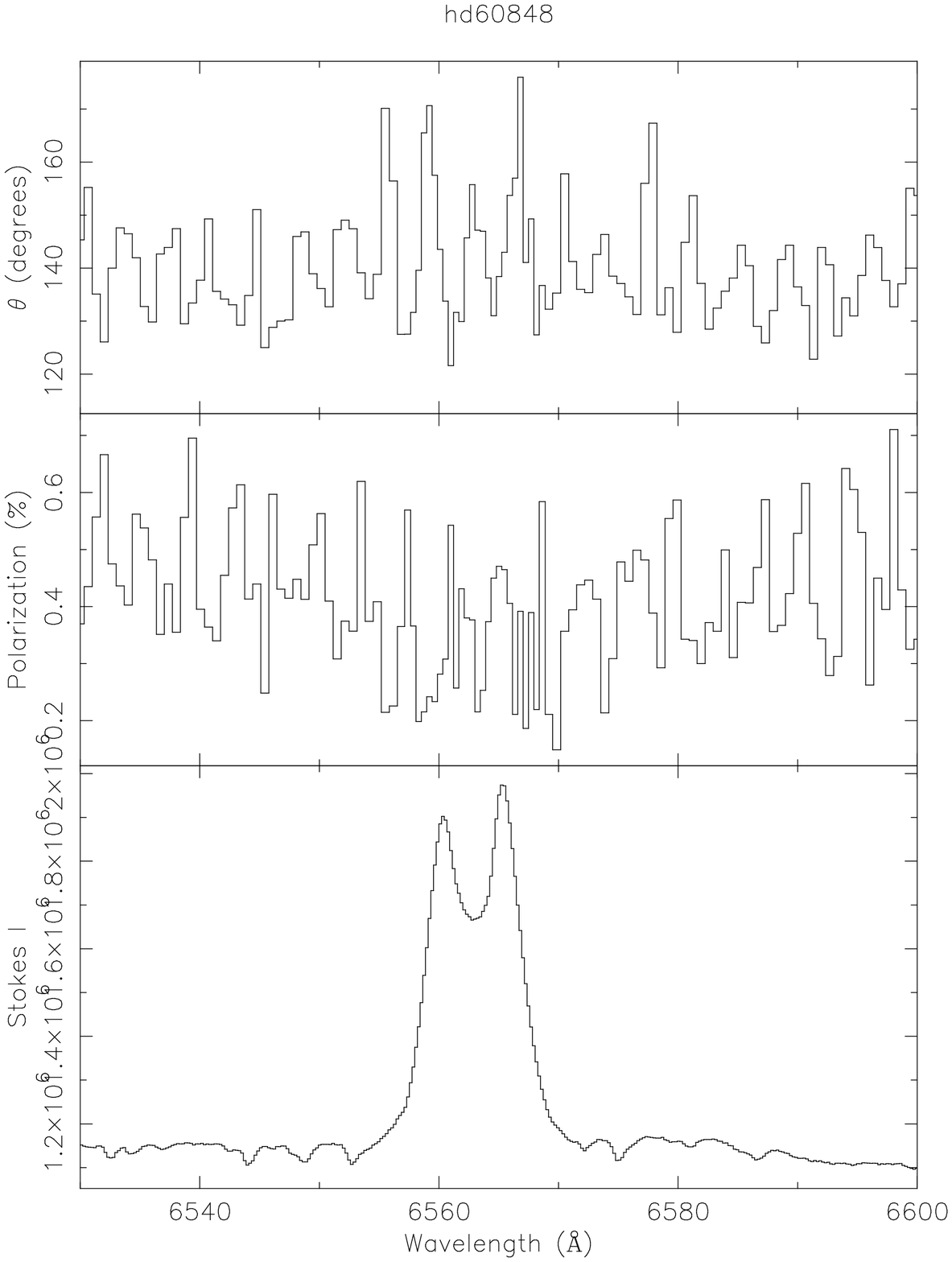}
\epsfxsize=0.25\textwidth\epsfbox{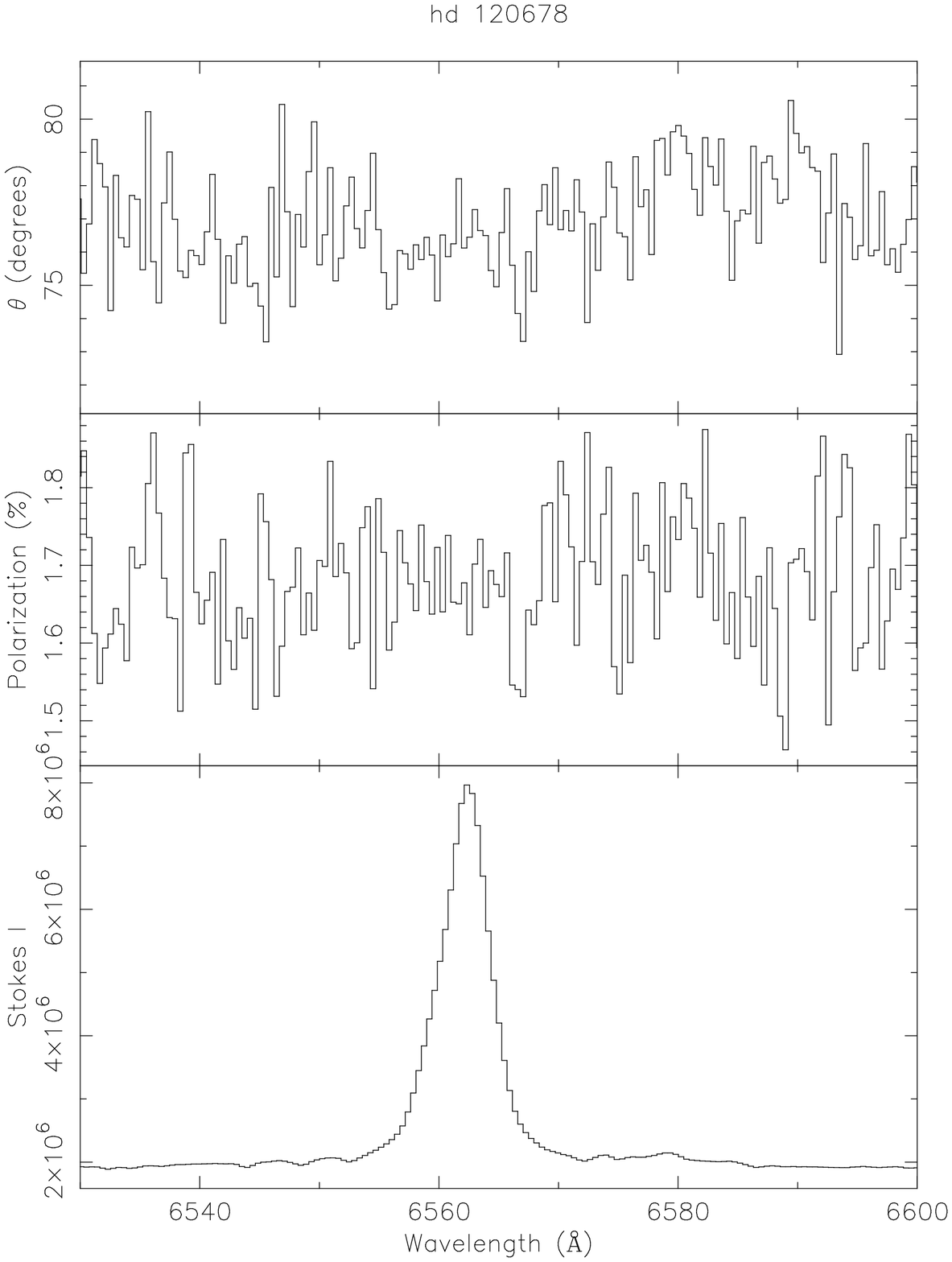}

}
\mbox{
\epsfxsize=0.25\textwidth\epsfbox{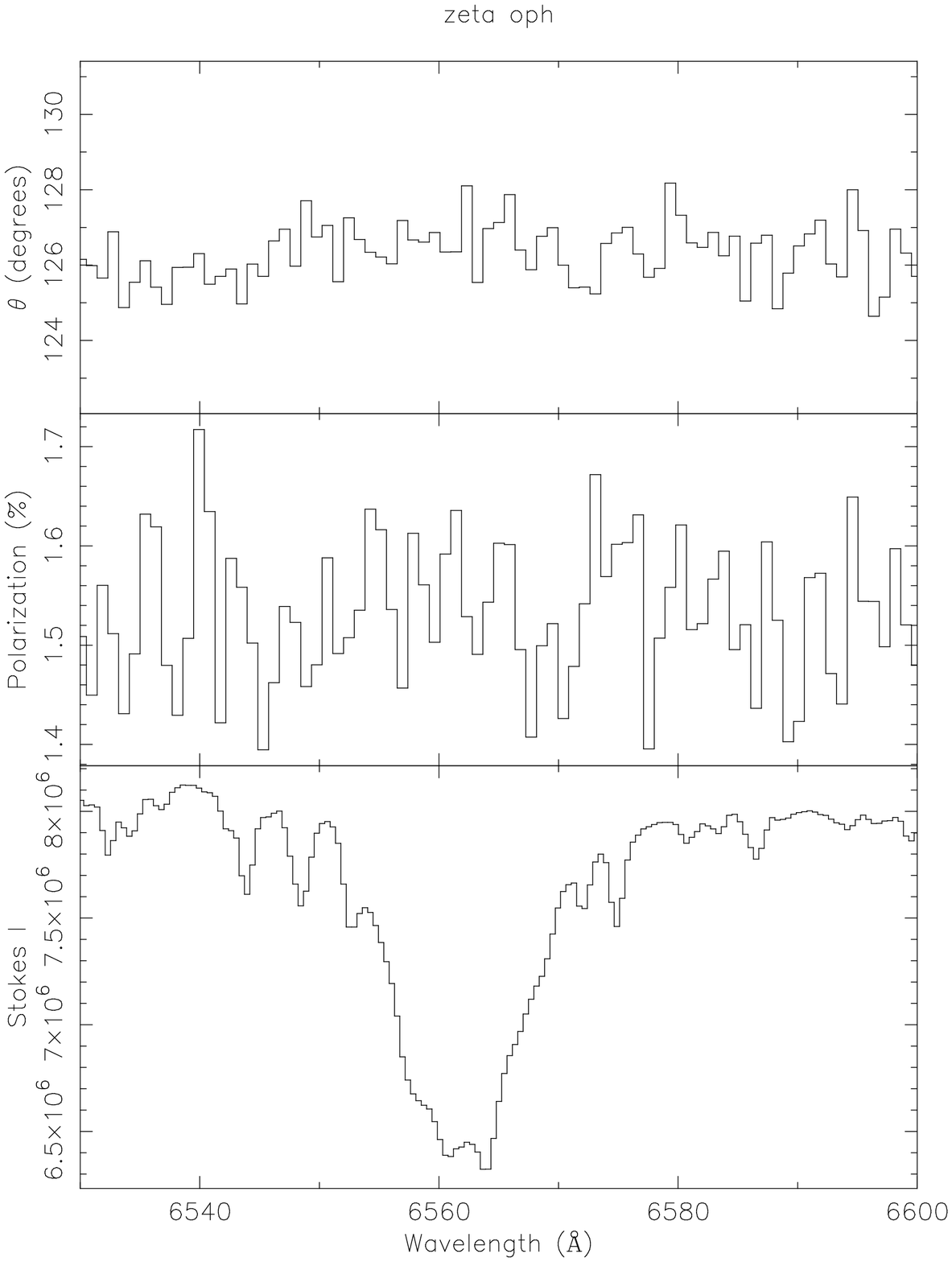}
\epsfxsize=0.25\textwidth\epsfbox{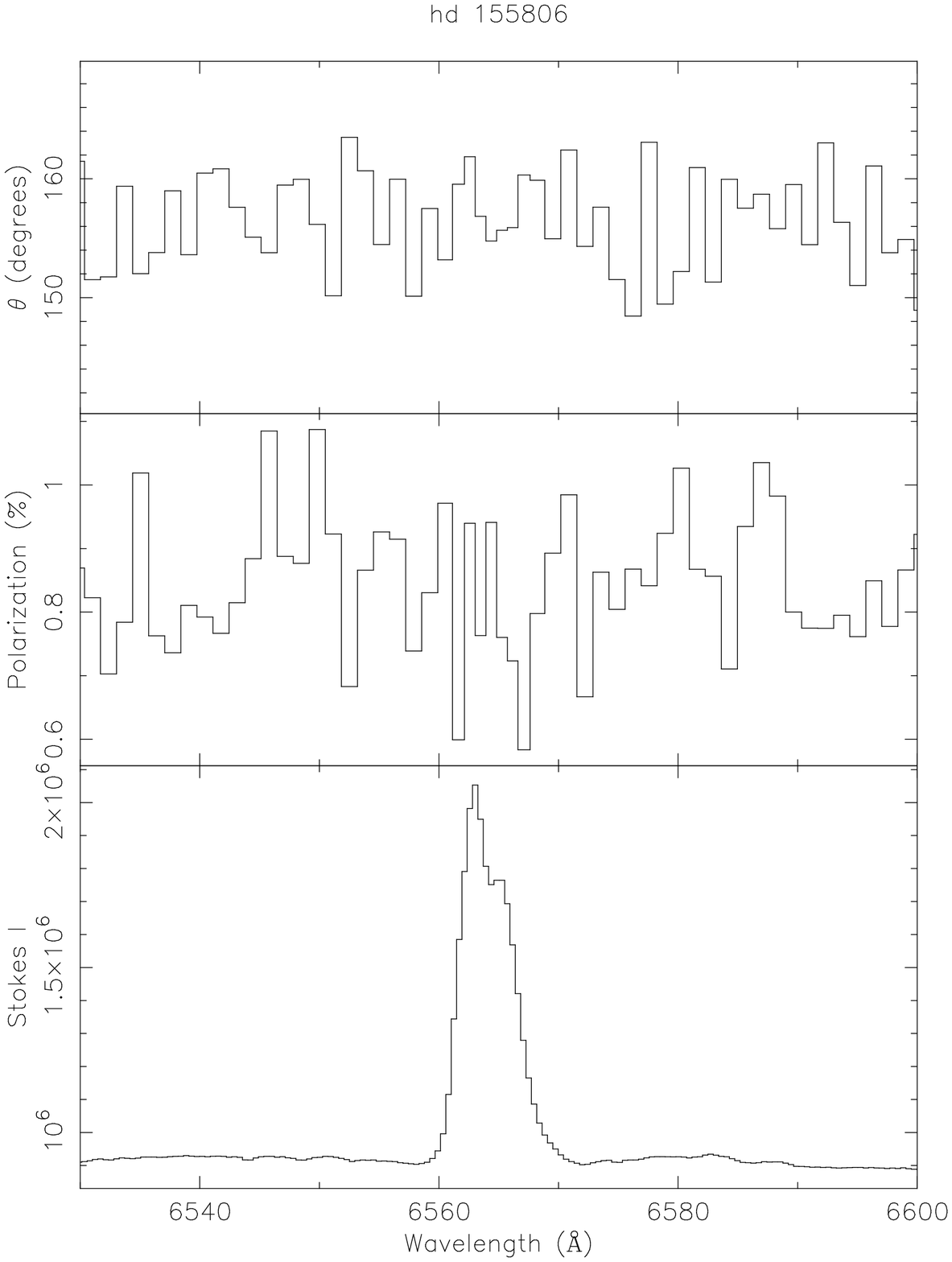}
}
\caption{
Polarization spectra of the Group {\sc iii} Oe stars. 
\label{f_oe}
}
\end{figure*}

\begin{figure}
\centerline{\psfig{file=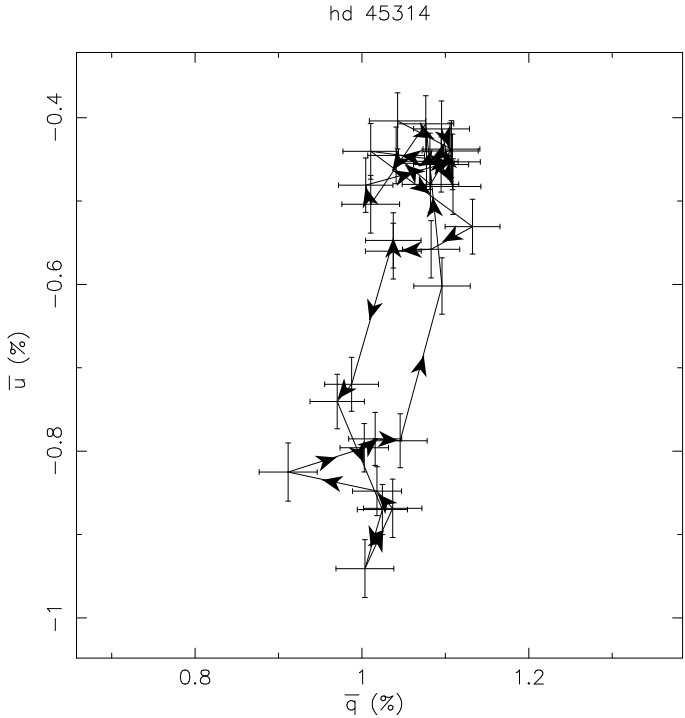, width=9cm }}
\caption{$QU$ diagram of the Group {\sc iii} Oe star HD\,45314. Note that the normalized Stokes 
parameters $u = U/I$ and $q = Q/I$ are depicted. The continuum polarization appears on the 
top of the diagram at $u$ $\sim$$-$0.45\%. The H$\alpha$ line excursion is shown (with an arrow) 
downwards to a value of $u$ as low as $\sim$$-$0.9\%.}
\label{f_qu_45314}
\end{figure}

The spectropolarimetry data for the six Oe stars are depicted in Fig.~\ref{f_oe}.
The Stokes I normal intensity data show a variety of line profiles, involving 
absorption, single, and double-peaked emission lines. HD\,45314 shows 
a clear polarization variation across the Stokes I emission profile that is 
as broad as that of the emission line itself. We can safely consider this effect 
to be the result of the classical ``depolarization'' line effect 
(see the extensive discussion in Vink et al. 2002). 
In such a case, one would expect to observe a change in \%Pol 
across the line that follows the shape of the intensity profile, but with the 
PA remaining constant. However as a result of the vector addition of 
interstellar polarization, it may be observable in the PA instead.
Indeed, although the line effect is present in both the 
polarization percentage (middle panel) and the PA (upper panel), the variation in the PA is not intrinsic 
to the star but a result of intervening ISP, which simply shifts the datapoints across the $QU$ plane. 
This conclusion is confirmed and explained when we alternatively plot the same data in a 
$QU$ diagram (see Fig.\ref{f_qu_45314}). In this representation the continuum polarization 
appears at the top of the diagram, corresponding to $u$ $\simeq$$-$0.45\%, whilst 
the H$\alpha$ line makes a linear excursion downwards (with the arrow denoting 
the wavelength direction) to a value of $u$ of $\simeq$$-$0.9\%. In other words, the intrinisic polarization
of HD\,45314 is of order 0.45\% at an intrinsic PA of $\simeq$135\degree. 

The other five Oe stars are not subject to any polarization line effects across H$\alpha$,  
although their Stokes I profiles show a variety of shapes. This is an intriguing result, esp. 
when one considers the fact that most $v$sin$i$ values are large (see Table 2) and double 
peaked emission is seen in non-line-effect objects like HD\,60848 and HD\,155806, which 
have strong Stokes I H$\alpha$ emission lines.
In the past, double-peaked emission lines in Stokes I intensity 
have often been interpreted as signalling circumstellar disks, but here we are reminded
by the fact that double-peaked profiles can for example also be produced in an expanding shell that is spherically symmetric
and would, irrespective of the inclination of the stellar rotation axis, not produce any linear polarization.

With respect to the continuum polarization measurements listed in Table~1, we 
count three objects for which we trust the ISP estimates. 
For the case of HD\,46485 the measured PA appears different from prior
measurements, as well as from the ISP. For HD\,120678 and HD\,155806 there is no notable 
difference between our PA measurements and earlier literature, nor do we note any 
variation with respect to their ISP estimates. We can safely assume these two objects 
are unpolarized. For the case of HD\,60848 our PA measurement of 136\degree\ deviates notably 
from a previous measurement of 56\degree\ but as the level of polarization is low, the error in the PA is 
quite large. The continuum PA measurement of 127\degree\ 
for $\zeta$Oph is consistent with the previous measurement, however the continuum 
polarization percentage has dropped to 0.6\%, compared to 1.4\% on a previous occasion, 
leaving open the possibility that this object is intrinsically polarized but that the wind emission 
at the epoch of observation was too weak to produce a line effect.

We summarize the Oe star polarization results in the following way. 
Out of the sample of six Oe stars, we can state that at least two of them are upolarized. 
If one were willing to accept time variability or ISP corrected polarization results, the remaining four 
Oe stars from our sample may be intrinsically polarized. 
However, when we do restrict ourselves to the original criteria for detecting intrinsically polarized O-type stars, 
only accepting objects that show polarization changes across spectral lines as sufficient evidence for 
intrinsic polarization, we find that only one out of six has a line effect. 
This fraction is statistically lower than that in classical Be stars, which have an 
incidence of line polarization effects of $\sim$55\% (Poeckert \& Marlborough 1976, Vink et al. 2002).  
The implications of this result are discussed in Sect.~\ref{s_disc}. 

\subsection{Onfp stars}
\label{s_onfp}

\begin{figure*}
\begin{center}
\mbox{
\epsfxsize=0.25\textwidth\epsfbox{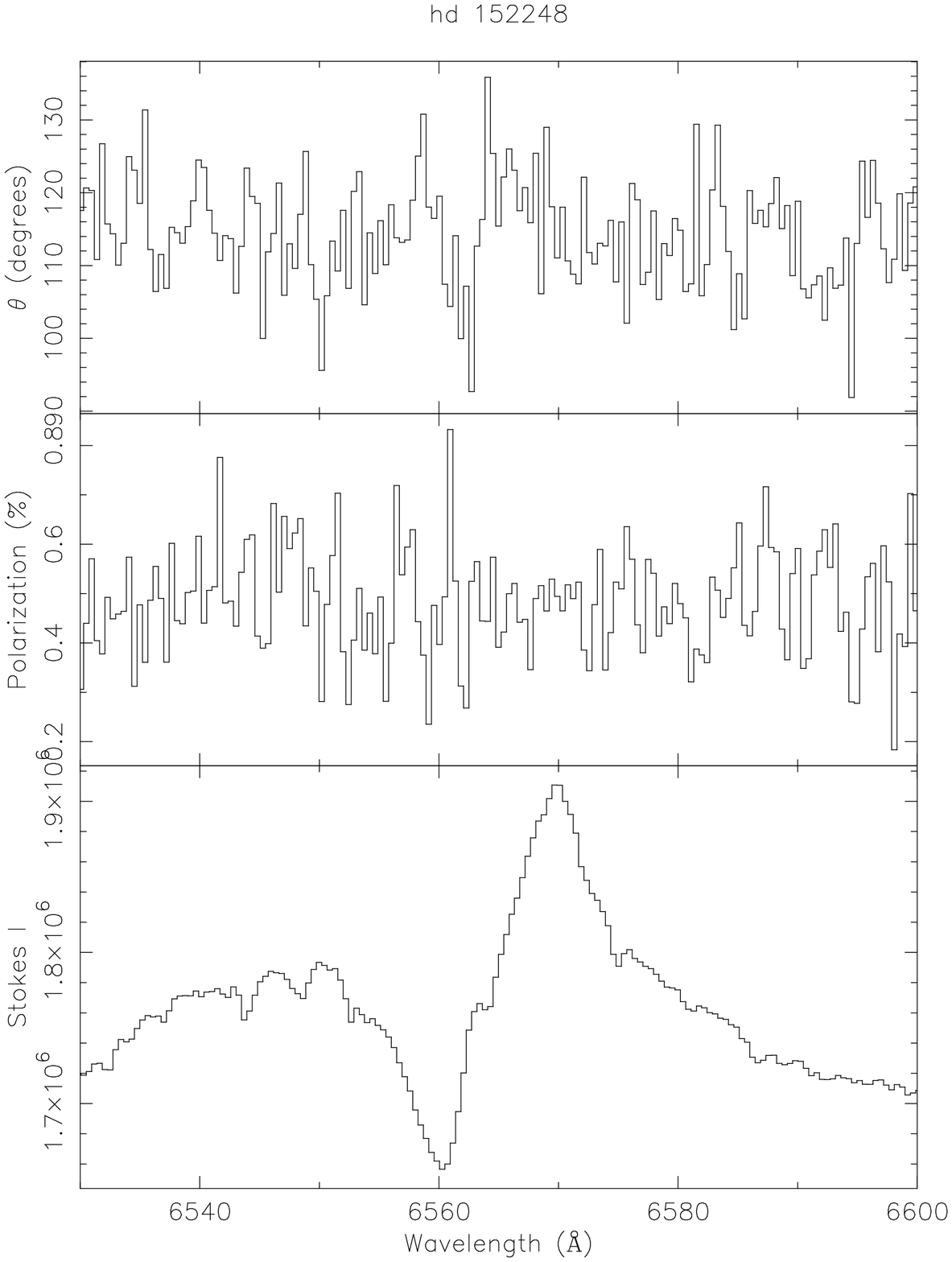}
\epsfxsize=0.25\textwidth\epsfbox{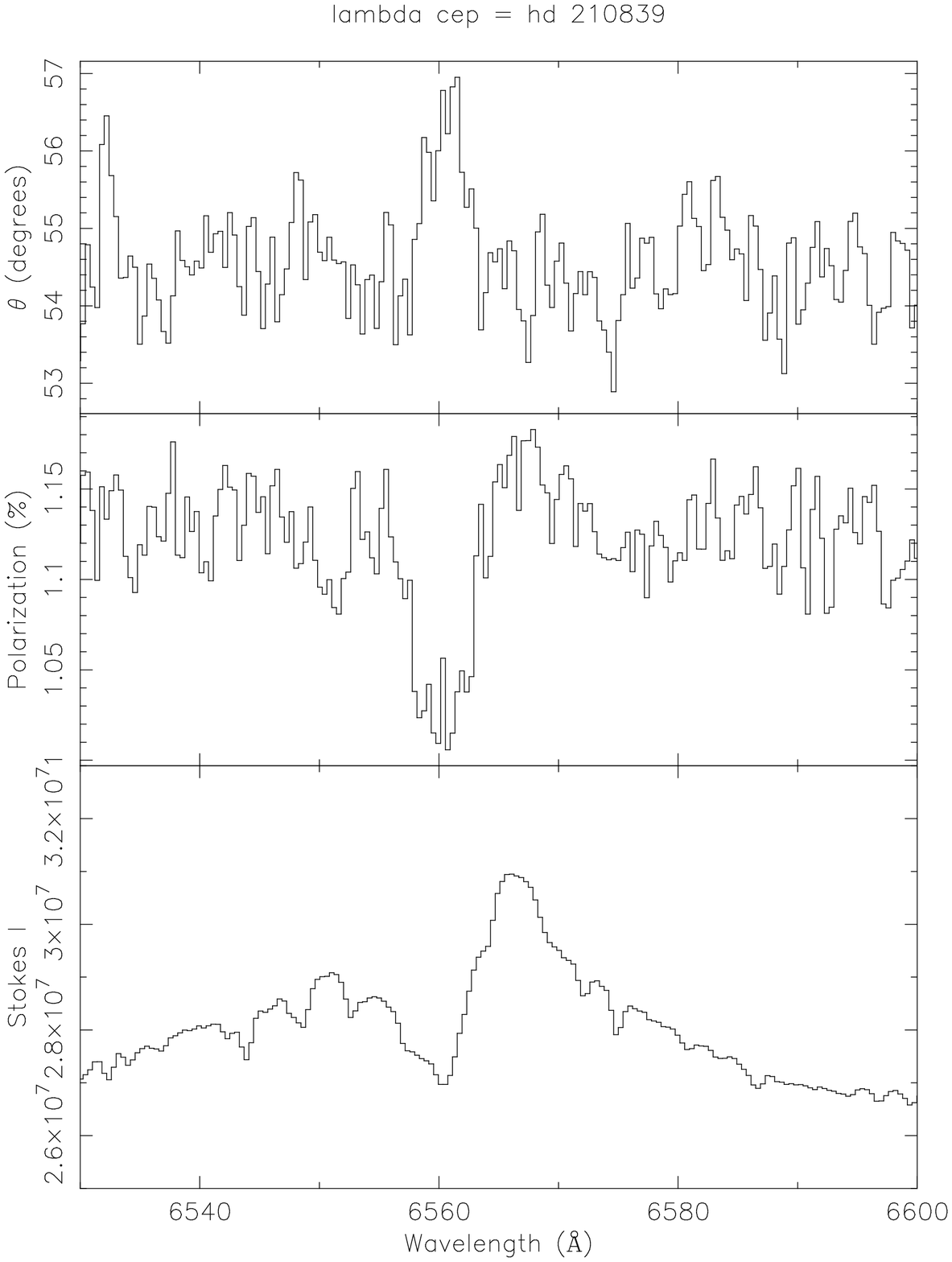}
\epsfxsize=0.25\textwidth\epsfbox{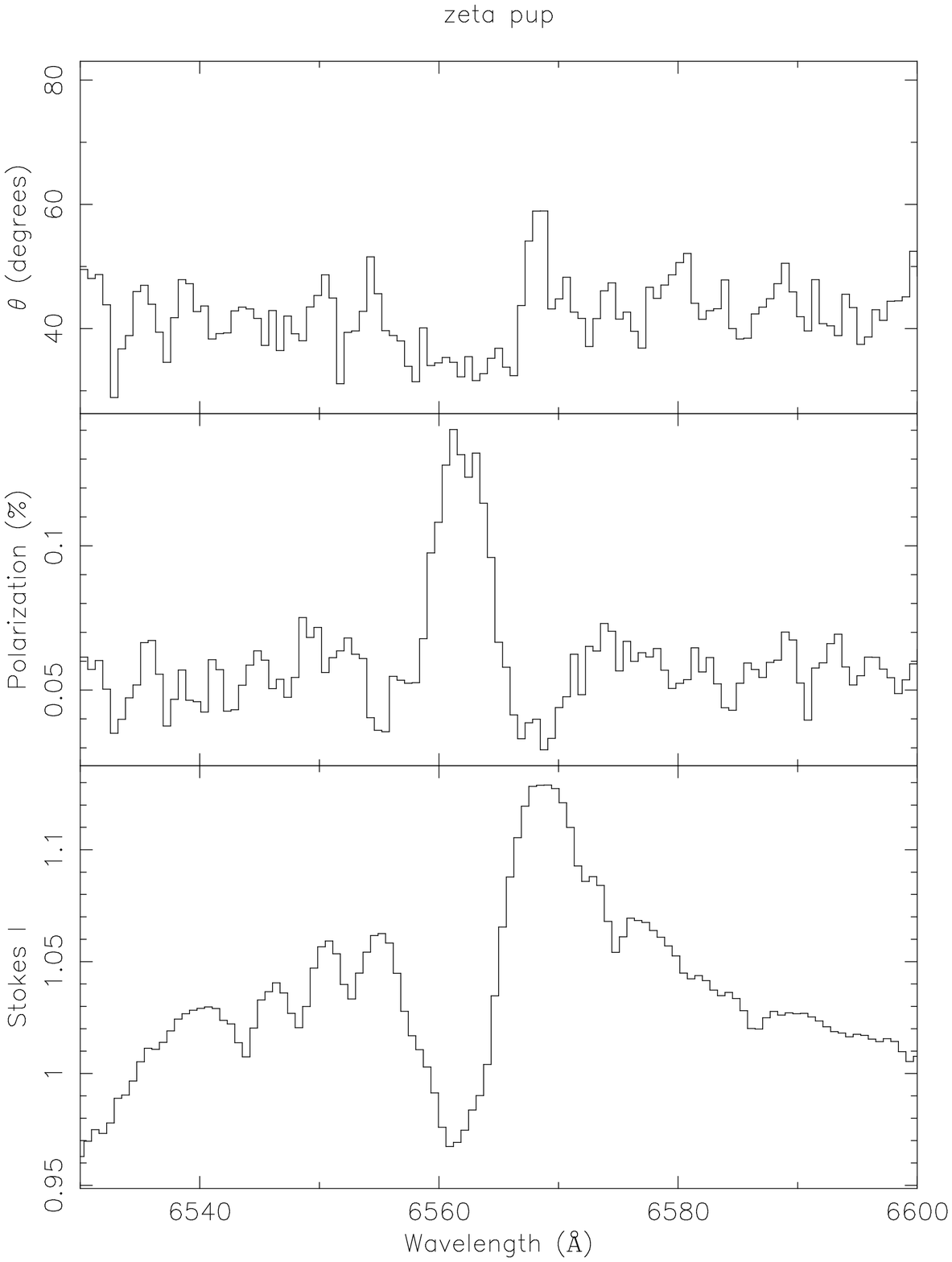}
\epsfxsize=0.25\textwidth\epsfbox{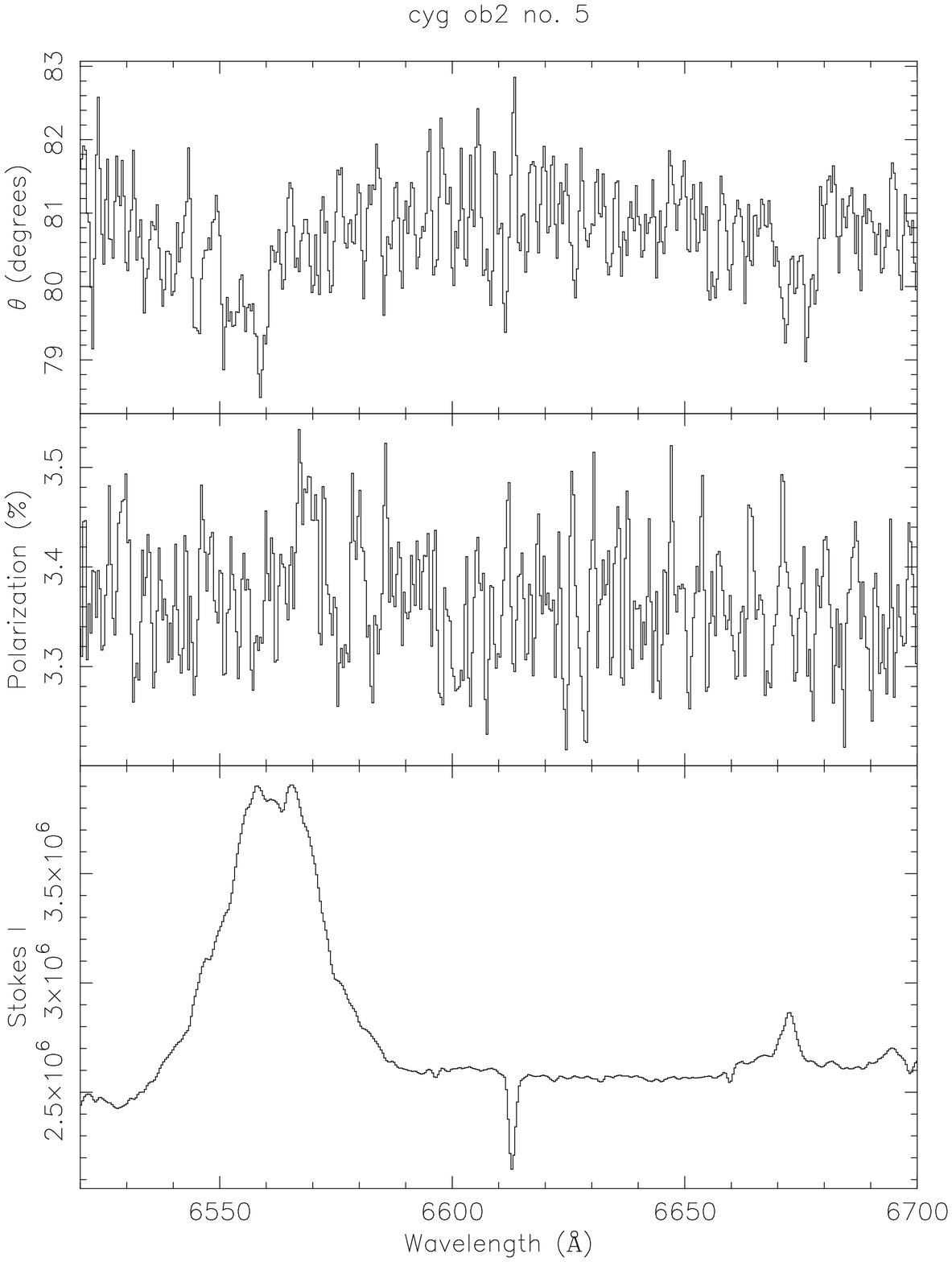}
}
\caption{Polarization spectra of the Group {\sc iv} Onfp stars HD\,152248, $\lambda$ Cep (HD\,210839), $\zeta$ Pup, and Cyg OB2 No.5 (from Harries et al. 2002).
Note that the wavelength range for Cyg OB2 No. 5 is wider than 
for the other stars.}
\label{f_onfp}
\end{center}
\end{figure*}

The Onfp stars are defined as objects in which He {\sc ii} 4686\AA\ shows centrally reversed 
emission (Walborn 1973), sometimes referred to as Oef stars (Conti \& Frost 1974).
In general these objects rotate rapidly, although there might be a variety of physical phenomena
that give rise to the strong He {\sc ii} emission, such as a strong stellar wind, or gas streams 
in close binaries. 

The spectropolarimetry data for the Onfp star HD\,152248 is presented in Fig.~\ref{f_onfp}, 
alongside 
data for three additional Onfp stars from Harries et al. (2002).  
HD\,152248 shows the kind of P Cygni Stokes I H$\alpha$ profile that are reminiscent 
of the more famous Onfp supergiants $\lambda$ Cep and $\zeta$ Pup.
The difference between them is that the very high signal-to-noise data for 
$\lambda$ Cep and $\zeta$ Pup show line effects (see Sect.~\ref{s_disc} for a discussion on the nature
of these line effects), whilst the lower signal-to-noise data of our HD\,152248 data 
only shows a tentative indication of a small variation in the PA, but the signal is 
not high enough to count HD\,152248 as a line-effect star at the present time.

\section{Discussion}
\label{s_disc}

In the following we place our results in the broader context 
of relevant characteristic information known about the four subgroups of O-type stars. 
This might provide constraints on formation models for massive stars, as well as 
the relevance of stellar rotation and magnetic fields with respect to 
disk formation around massive stars.

To help assessing the probability of finding a line effect in an individual
object or a subgroup of peculiar O-type stars, we quantify a representative wind density of our
objects, $\rho_{\rm wind}=\mdot/(R^2 \vinf)$ (column 9 of Table~\ref{t_params}). 
Note that we choose theoretical over empirical mass-loss rates (as determined from 
quantitative spectroscopy), as only 5 out of these 23 targets have empirical 
mass-loss rates available derived from line-blanketed models 
(see Repolust et al. 2004, Martins et al. 2005b, Mokiem et al. 2005, 
Marcolino et al. 2009). 
Furthermore, the rates derived from different spectral ranges 
appear to be highly uncertain. For instance, for $\zeta$ Oph, Marcolino et al. (2009) provide 
a mass-loss rate as low as log$(\Mdot)$ = $-$8.80 from the ultraviolet, whilst Mokiem et al. (2005) 
give log($\Mdot$) = $-$6.84, determined from H$\alpha$. The latter value is in very good agreement with 
the theoretical value of log($\mdot$) = $-$6.83 (see column 8). In other words, the use of theoretical rates 
is not only preferred for practical reasons. 
We note that if the terminal velocity is not available, we 
multiplied the escape velocity by the canonical factor 2.6 for stars on the hot side of the bi-stability jump 
(Lamers et al. 1995).

As can be noted from column 9 of Table~\ref{t_params}, the average wind density per subgroup 
does not appear to be very different. Interestingly, the line-effect Oe star HD\,45314
is actually an object with one of the smallest wind densities given in column 9. 
We can thus conclude that the average wind density 
does not play a major role in determining whether or not a certain 
subgroup of O stars is subject to linear polarization. Rather, it is predominately 
determined by the equator/pole density ratio, 
$\rho_{\rm eq}/\rho_{\rm pole}$ which is required to be larger than 1.25, in order to achieve 
the measurable amount of linear polarization of 0.1\% or more (see Sect.2).

\subsection{Group I: OVz and weak-wind stars}

Given that this subgroup may contain a number of young O-type stars, one might possibly expect that some could be 
surrounded by circumstellar material leftover from the 
star formation process -- in a manner analogues to pre-main sequence Herbig Ae/Be stars. 
For this reason, we searched the 2MASS catalogue for $JHK$ magnitudes, but we did not find 
OVz stars to have a large near-infrared excesses (of more than $\sim$ 0.5 mags), from which we conclude that it is unlikely 
that a large fraction of OVz and related objects be surrounded by large quantities of 
circumstellar dust. However, gaseous circumstellar disks could potentially be detected using
linear spectropolarimetry, if (i) these objects are sufficiently young, and (ii) if 
massive O stars form via disk accretion. 

A second reason why one might anticipate the presence of a gaseous 
circumstellar disk in these objects is that its most famous member 
$\Theta^1$ Ori C has a strong dipolar magnetic field (Donati et al. 2002, Wade et al. 2006) and probably
a magnetically confined wind leading to a pile-up of gaseous material around 
the magnetic equator like in the Bp star $\sigma$ Ori E 
(e.g. Babel \& Montmerle 1997, ud-Doula \& Owocki 2002).
 
\begin{table*}
\caption{Stellar parameters for the four groups of peculiar O-type stars.}
\label{t_params}
\begin{tabular}{lcccccccccc}
\hline
Name & log($L/\lsun$) & $\teff$ & log$g$ & $M$       & $R$       & $\vinf$ & log$\mdot$ & $\rho_{\rm wind}$ & $v$sin$i$ \\
     &                & (kK)    &        & $(\msun)$ & $(\rsun)$ & (km/s)  & ($\msunyr$) & ($10^{-13}$ \msunyr/\kms${\rsun^2}$) & (km/s)\\
\hline
\\
Group I~~{\it OVz and weak wind dwarfs}:\\
\\
\\
$\Theta^1$ Ori C    &  5.13  & 39  & 4.1  & 30   & 8  & 580 & $-$5.7 & 530 & 24\\
$\Theta^2$ Ori A    & 4.62  & 30.5 & 3.92 & 17 & 7  & 2500$^1$    & $-$7.59 & 1.9 &     \\
HD 42088 & 5.23  & 38   &   4.0 & 32        & 10 & 1900 & $-$6.19 & 37 & 60 \\
HD 54662           & 5.2   & 36.8 & 3.92 & 29 & 10  & 2456& $-$6.40 & 17 &  73   \\
HD 152590          & 5.00  & 34.4 & 3.92 & 24 & 9  & 1785& $-$6.64 & 23 & 60    \\
HD 93128           &  5.66 & 46.5 & 4.00 & 40 & 10 & 3210$^1$    & $-$5.53 & 100 & 100 \\
HD 93129B          &  5.86 & 44.0 & 3.92 & 52 & 13 & 3210$^1$    & $-$5.25 & 110 &      \\
CPD$-$58\degree 2611 &  5.3  & 38.2 & 3.92 & 32 & 10 & 2870$^1$    & $-$6.23 & 21  &      \\
CPD$-$58\degree 2620 &  5.2  & 36.8 & 3.92 & 29 & 24 & 1760$^1$    & $-$6.43 & 15 & 110     \\
\\
\hline
\\
Group II~~{\it Of?p}:\\
\\
\\
HD 148937     &     5.75  & 41.0  & 4.0  & $\sim$55   & 15 & 2285 & $-$5.45 & 71 & 76 \\
HD 108        &     5.40  & 37.0  & 3.75 & $\sim$35   & 12 & 1960 & $-$5.97 & 37 & 78 \\
HD 191612     &     5.4   & 35    & 3.5  & $\sim$35   & 15 & 2450$^1$  & $-$6.14 & 17 &   \\
\\
\hline
\\
Group III~~{\it Oe stars}:\\
\\
\\
HD 39680  & 5.5  & 40   &       & 45        & 11   & 1675 & $-$5.69 & 90 & 218   \\
HD 45314 & 4.72  & 31.5 & 3.92  & 18        & 8  & 2410$^1$     & $-$7.37 & 3.0  & 285 \\
HD 46485  & 5.1  & 35.5 & 3.92  & 27      & 9  & 1800 & $-$6.44 & 25 & &      \\
HD 60848  & 4.9  & 33.4 & 3.92  & 22        & 8.5  & 1765 & $-$6.84 & 12 & 240   \\
HD 120678 & 4.9  & 33.4 & 3.92  & 22        & 8.5  & 2580$^1$ & $-$7.02 & 5.5 &     \\
Zeta Oph & 4.88  & 32.1 & 3.62  & 20        & 9  & 1505 & $-$6.83 & 12 & 372   \\ 
HD 155806 & 4.9  & 33.4 & 3.92  & 22        & 8.5  & 2460 & $-$7.02 & 5.5 & 91   \\
\\
\hline
\\
Group IV~~{\it Onfp}:\\
\\
\\
HD 152248A & 5.69 & 33.3 & 3.4  & 41   & 21.1 & 2420 & $-$6.10 & 16 & 159    \\
HD 210839 & 5.83 & 36.0 & 3.55  & 62   & 21.1 & 2250 & $-$5.47 & 33 & 200 \\
Zeta Pup  & 5.90 & 39.0 & 3.60  & 54   & 19.4 & 2250 & $-$5.24 & 68 & 219  \\
Cyg OB2 No.5 & 5.69 & 33.3 & 3.4& 41   & 21.1 & 2240$^1$ & $-$5.64 & 27 &     \\
\\
\hline
\end{tabular}
\\
\noindent
The stellar parameters are taken from the following papers:
Repolust et al. (2004), Martins et al. (2005b), Puls et al. (2006). For all stars without recent line-blanketed 
models, the calibrations of Martins et al. (2005a) are used. The mass-loss rates are from Monte Carlo 
radiation-driven wind models (Vink et al. 2000). The wind density is given as 
$\rho_{\rm wind}=\mdot/(R^2 \vinf)$. 
Where spectral subtypes are not provided, we interpolate. $v$sin$i$ are mostly taken from Howarth et al. (1997).\\
\\
$^1$ Note that the terminal velocity is calculated by multiplication of the escape velocity 
by the canonical factor of 2.6 (Lamers et al. 1995).\\
\end{table*}

We can summarize the results of our linear H$\alpha$ line spectropolarimetry of the 
nine Group {\sc i} OVz and related stars with the simple statement that none of 
them shows strong evidence for linear polarization, and we consider a number of different 
reasons why this might be so.

Perhaps the simplest explanation would be that the absorptive character of the 
Stokes I H$\alpha$ profile prevents our technique from being sensitive enough. 
However, it would not explain the non-detection in the Dec 
1995 data of $\Theta^1$ Ori C when H$\alpha$ was seen in emission (see Fig.~\ref{f_theta}). 
Even if our linear polarimetry tool fails to detect disks when H$\alpha$ is mostly photospheric, and 
even if accretion disks are absent, it is still somewhat puzzling that even for
$\Theta^1$ Ori C in which the presence of a magnetically confined wind seems well-established, we 
do not detect any evidence for flattened circumstellar material in 
our OVz linear polarimetry data. 

It is possible that the circumstellar disk structures are disrupted by the magnetic pressure of the 
central object and that gaseous material is only present at larger radii, with
an inner hole present closer to the star, where the number of free electrons would be lower than for 
normal O stars, limiting the level of linear continuum polarization. 
In other words, the absence of line effects in $\Theta^1$ Ori C might actually be consistent with a 
magnetically confined wind scenario, where the magnetic pressure is too high close to the surface, 
creating an inner disk hole. It would be interesting to provide further constraints
on the size of such a hole and this might yet again be achieved using line polarimetry. 
Differences in the shapes of {\it intrinsic} line polarimetry (in contrast to line depolarization where the 
{\it continuum} is polarized) involving either single or double 
loops in the $QU$ plane have the potential to constrain the size of disk inner holes 
(Vink et al. 2005a). This could work in case the line under consideration is in emission, which
occurs quite regularly for objects like $\Theta^1$ Ori C (see Fig.~\ref{f_theta}) as well as 
the Bp star $\sigma$ Ori E. 
It will be interesting to see whether this technique could be applied to the bulk of the Group {\sc i} 
OVz and related objects. 

We note that Smith \& Fullerton (2005) re-examined the velocity behaviour of UV wind 
lines in $\Theta^1$ Ori C and found that in contrast to $\sigma$ Ori E, the UV profiles of $\Theta^1$ Ori C
show surprisingly little evidence for the presence of a pile-up of material, suggesting that despite the 
strong dipolar magnetic field of $\Theta^1$ Ori C, it is too early to conclude the 
star is simply a massive analog of Bp stars.
To continue the discussion along this line of reasoning, the small variations in the 
continuum polarization of $\Theta^2$ Ori A may be related to the complex gas 
flows around the object. 
We recommend the Group {\sc i} OVz and related objects 
be monitored simultaneously in linear and circular polarization to obtain a more comprehensive picture of the 
geometry of the magnetic field in conjunction with that of the circumstellar material.

\subsection{Of?p stars}

The Of?p stars appear to be relatively slow rotating stars (see Table 2) 
that show dramatic periodic spectral variability (e.g. Naze et al. 2008).  
Contrary to the OVz object $\Theta^2$ Ori A, 
the PA of the Of?p star HD\,148937 is constant with time, although  
it may be subject to variations in the level of linear polarization. 
Out of the 20 supergiants studied by Harries et al. (2002), the one 
Of?p star that was studied, HD\,108, was one of only five objects to show a line effect. 
The character of the HD\,108 line effect 
is consistent with classical depolarization (see Fig.~\ref{f_of?p})

The third known Galactic Of?p star, HD\,191612, has not yet been observed in linear spectropolarimetry mode, but interestingly 
it is one out of only three O-type stars\footnote{The other two O-type stars with confirmed magnetic fields are $\Theta^1$ Ori C 
(Donati et al. 2002, Wade et al. 2006), and $\zeta$ Ori (Bouret et al. 2008b)} in which a magnetic field has been discovered through 
circular spectropolarimetry (Donati et al. 2006, Howarth et al. 2007). 
Similarly to the comments made above with respect to the Group {\sc i} objects like $\Theta^1$ Ori C, it
would be most informative to monitor the Of?p stars simultaneously in linear and circular polarized light. 
The fact that there is no evidence for rapid rotation in any of the known Of?p stars (see Table \ref{t_params})
might be consistent with a scenario of magnetic braking for these objects.

\subsection{Oe stars}

The Oe stars have been suggested to be related to classical Be stars.
The definition of classical Be stars is that they are near-main sequence 
B stars (of luminosity class {\sc iii - v}) in which the hydrogen Balmer lines are or have been in emission. 
The emission lines often show blue- and red-shifted peaks, consistent with the presence 
of a rotating circumstellar disk (Struve 1931), and this disk model was confirmed decades 
later through both polarization (e.g. Poeckert \& Marlborough 1976, Wood et al. 1997) and 
interferometry measurements (Quirrenbach et al. 1997). The situation for the Oe stars
is however less clear. On the basis of the Stokes I spectrum, 
Conti \& Leep (1974) suggested these objects simply represent the 
higher mass analogues of the classical Be stars, but in more recent work of 
Negueruela et al. (2004) it was found that many of these objects were in fact 
spectroscopically classified too early, with the possibility that the more massive 
counterparts to classical Be stars, i.e. classical Oe stars, simply do not exist.

The relevance of the general absence of Oe stars is that it could provide insight 
into the physical mechanism underlying the Be phenomenon, as some theoretical
models for explaining the Be phenomenon predict disks to be present across the 
entire spectral range, whilst other models predict the presence of disks
solely at certain spectral types, corresponding to specific effective temperatures, and possibly 
restricted to specific stellar masses and/or ages (e.g. Fabregat \& Torejon 2000, McSwain \& Gies 2005, Martayan et al. 2006).
Reasons for an absence of the hotter analogues of classical Be stars 
could involve the more intense ionizing radiation or stronger radiatively-driven wind
of the hotter O stars, weaker wind braking for B stars (e.g. Maeder \& Meynet 2000), or 
special circumstances, such as wind bi-stability (Pauldrach \& Puls 1990, Vink et al. 1999) to 
explain the presence of circumstellar disks exclusively amongst objects that fall in the B-star spectral range. 

Even though Negueruela et al. (2004) found some of the Oe stars to be classified ``too early'', 
it is still true that some of them fall within in the O-star spectral range, and as a population
they are rapid rotators (see the large $v$sin$i$ values in Table \ref{t_params}). The relevant 
question is now whether these objects are embedded in geometrically thin disks, analogous to those present 
around classical Be stars. When we take our result of one out of six line depolarization 
effects in the Group {\sc iii} Oe stars at face value and compare this incidence to that amongst 
the classical Be stars, whilst properly accounting for the different sample sizes, we 
find that the chance the line effect data of the Be and Oe samples are drawn 
from the same parent sample is negligible, at 95\% confidence.
In other words, contrary to common belief, there is no evidence 
that early-type Oe stars such as HD 155806 are embedded in circumstellar disks. 

We subscribe to the view of Negueruela et al. (2004) that the 
group of Oe stars, and especially those with spectral types earlier than O9.5, deserve more 
detailed investigations to provide better constraints on the physical origin of the Be phenomenon.

\subsection{Onfp stars}

With respect to the incidence of line polarization effects for the Onfp stars, 
the situation appears to be different from that in the Oe stars discussed above.
Out of the four Onfp stars studied spectropolarimetrically\footnote{Note that we have now counted 
$\zeta$ Pup as an Onfp star. According to Walborn's spectra the object is not Onfp, but Conti noticed 
the characteristic He {\sc ii} 4686\AA\ emission in his Oef notation, indicating variability.}, at least three 
of them show evidence for line effects across H$\alpha$.

We should note that the physical
phenomena that give rise to the peculiar He {\sc ii} emission might
be due to differing physical origins. Cyg OB2 No. 5 is classified as O7 Ianfp, but 
this object does not display the characteristic Onfp He {\sc ii} 4686\AA\ profile. 
Instead, it is a peculiar and variable interacting binary. 
The He {\sc ii} 4686\AA\ emission 
in HD\,152248 may also be due to gas streams in a close binary. This object
was recently studied by Mayer et al. (2008) and the stellar mass of $\simeq$30 \Msun\ in that paper 
is probably more accurate than the stellar mass listed in Table 2 (as it is simply derived from the binary motion).
However, we find the H$\alpha$ shape in HD\,152248 very reminiscent of that in $\lambda$ Cep and $\zeta$ Pup. 

With respect to the nature of the line effects in Onfp stars we cannot draw any definitive conclusions, yet 
we make the following remarks. 
The fact that the polarization changes in $\lambda$ Cep occur predominantly across blue-shifted absorption 
might indicate that it is not the line depolarization effect that is at work here. Instead, these polarization 
profiles seem to be more reminiscent 
of the ``McLean'' effect (McLean 1979) in Herbig Ae/Be stars (Vink et al. 2002) or of optically pumped gas 
(Kuhn et al. 2007). However, the most convincing modelling so far suggests that these line effects are 
the result of the star's rapid rotation, which causes an asymmetry in velocity space (Harries 2000). 

Given the small sample of Onfp stars studied so far and the complexities involved in 
the line polarization profiles of the Group {\sc iv} Onfp stars, it would 
be premature to conclude that the current sample of line polarization effects need necessarily 
indicate the presence of disks around Onfp stars. 
It should also be noted that geometrical effects other than disks, such as
wind clumping (see Davies et al. 2007 and references therein) might also give rise 
to the observed levels of linear polarization. Indeed, it may also present an alternative explanation for 
the peculiar Stokes $I$ spectrum of $\zeta$ Pup (Bouret et al. 2008a).
Nonetheless, similarly to Group {\sc iii} Oe stars, Group {\sc iv} Onfp stars appear to be 
rapid rotators (see Table \ref{t_params}), with the difference being that the incidence of line effects in Onfp stars 
appears to be consistent with that expected for objects surrounded by disks with random orientations, although this is based on 
extremely low number statistics. 

Given their rapid rotation and the range of their intrinsic luminosities (log($L/\lsun$) $\simeq$5.7 - 5.8; see Table 2), it is 
tempting to speculate the Galactic Onfp stars are the precursors of the Galactic B[e] supergiants, 
which are also believed to be rotating rapidly but for which the evidence of equatorial disks seems better 
established (Zickgraf et al. 1985).
Interestingly, the B[e] phenomenon has been found to extend to significantly lower luminosities in 
the Large Magellanic Cloud (Gummersbach et al. 1995) and it will be informative to see 
if the Onfp phenomenon is also observed towards lower luminosities in the LMC (see Walborn 2008).

Finally, it will be crucial to 
enlarge the database of linear spectropolarimetry data for these Group {\sc iv} Onfp stars, as they 
appear to be the most promising subgroup of 
O-type stars of being a significant disk producing massive star population, with potential implications for 
constraining progenitor models for gamma-ray bursts and other exotic phenomena that may be related to rapid 
rotation.

\section{Summary}
\label{s_sum}

We have presented linear spectropolarimetry data on a sample of 18 peculiar 
O-type stars, supplementing an earlier sample of 20 mostly normal O stars of Harries et al. (2002),
yielding a total of 38 O-type stars. On the basis of the spectral Stokes I peculiarities we 
divided these peculiar objects into four separate groups and studied their 
linear Stokes $QU$ polarization properties.\\

\noindent
Group I~~~included the suspected young zero-age main sequence OVz stars and related objects.
For this group we can summarize the results with the simple statement 
that none of them shows evidence for significant amounts of linear polarization.
It is not inconceivable that circumstellar disks would be disrupted by the magnetic pressure of the 
central object, and the absence of polarization line effects might be considered 
consistent with a magnetic scenario, where the magnetic pressure could be responsible for an inner hole, which
limits the number of free electrons close to the stellar surface and the associated level of 
linear continuum polarization.

We also suggested that the measured variations in the 
continuum polarization of objects like $\Theta^2$ Ori A may 
be related to the complex gas flows associated with its binary nature or 
the formation of magnetically confined disks and we recommend Group {\sc i} objects 
be simultaneously monitored in linear and circular polarized light.\\

\noindent
Group II~~~included the spectroscopically variable Of?p stars, which seem 
to comprise a group of slowly spinning stars that are subject to 
dramatic periodic spectral variability. 

Contrary to the Group {\sc i} stars like $\Theta^2$ Ori A, 
the PA of the Of?p object HD\,148937 is constant with time, although  
it may be subject to variations in the level of linear polarization (a 2 $\sigma$ effect). 
Interestingly, a second member of the group HD\,108 was one of the few O-type stars 
in Harries et al. that showed a line effect. 
The third member HD\,191612 was not observed in linear spectropolarimetry mode, but interestingly 
it was recently found to be magnetic, possibly providing a link to some of the younger magnetic stars 
from group {\sc i} such as $\Theta^1$ Ori C. 

Given the high incidence of both linear and circular polarization plus the fact that 
there is no evidence for rapid rotation in any of the Of?p stars 
(see Table \ref{t_params}), leads us to speculate that it is the presence of a 
magnetic field that is the underlying reason for this peculiar spectral classification.\\

\noindent
Group III~~~includes Oe stars, which have been suggested to be the more massive counterparts to 
classical Be stars. However, when we take our linear H$\alpha$ polarimetry results at face value 
and compare the low incidence (one out of six) with that of the classical Be stars (26 out of 44),
thereby properly accounting for the different sample sizes, we conclude that the chance that the 
line effect Oe and Be
samples are drawn from the same parent sample is negligible at the 95\% confidence level.

In other words, there is as yet no evidence that stars with spectral types different from the B-type range possess 
circumstellar disks.\\

\noindent
Group IV~~~concern the rapidly rotating Onfp stars. 
Similarly to objects from group III, evidence for a disk hypothesis has yet to be found
for this subgroup.  
By contrast to the Oe stars, out of four Onfp stars studied, at least 
three show evidence of line polarization effects. 
Due to small number statistics it is not possible to conclude that our results imply 
the presence of disks, but interestingly the high incidence of 
line effect Onfp stars does appear to be in line with that expected for objects 
with random orientations (Poeckert \& Marlborough 1976, Vink et al. 2002). 

We speculated that the Galactic Onfp supergiants could be the precursors of B[e] supergiants, 
given their similar luminosity range (around log($L/\lsun)$ $\simeq$5.7-5.8), their rapid rotation, and possibly 
the presence of circumstellar outflowing disks in both object classes. 
It will be interesting to enlarge the current linear polarimetry dataset for group {\sc iv} Onfp 
objects, as these objects appear to be the most promising subgroup of O-type stars with line effects.\\

To conclude, we presented the first linear $QU$ polarization data for O-type stars that exhibit Stokes $I$
peculiarity. We showed that those objects which are peculiar in Stokes $I$ have a higher chance 
of showing $QU$ polarization. Future Stokes $V$ circular polarimetry studies of both normal
and peculiar O-type stars e.g. via the {\sc mimes} collaboration with CFHT/Espadons should 
teach us how the incidence of magnetic fields compares to the Stokes $I$ and Stokes $QU$ phenomenology described
here, as the rotational evolution of massive O stars is probably driven by a complex interplay between mass loss, rotation, 
and magnetic fields. This interplay needs to be constrained empirically and understood theoretically 
before we are able to explain {\it which} 
massive stars produce exotic phenomena such as long-duration gamma-ray bursts.

\begin{acknowledgements}

We wish to thank the anonymous referee for very constructive comments that have helped improve the paper, and the 
friendly staff at the AAT and WHT for their help during our 
observing runs. The allocation of time on the AAT and WHT was awarded by PATT, 
the UK allocation panel. This publication makes use of data products from the Two Micron All Sky Survey, 
which is a joint project of the University of Massachusetts and the Infrared Processing and Analysis 
Center/California Institute of Technology, funded by the National Aeronautics and Space Administration 
and the National Science Foundation.
This research has made use of the SIMBAD database, operated at CDS, Strasbourg, France.

\end{acknowledgements}

\end{document}